\documentclass[arxiv, preprint]{imsart}
\RequirePackage[authoryear]{natbib}
\RequirePackage[OT1]{fontenc}
\RequirePackage{amsthm,amsmath,amsfonts,graphicx,float,multirow,enumitem}
\RequirePackage[colorlinks,citecolor=blue,urlcolor=blue]{hyperref}
\usepackage{xr}
\usepackage[bottom]{footmisc}

\startlocaldefs
\numberwithin{equation}{section}
\theoremstyle{plain}
\newtheorem{thm}{Theorem}[section]
\newtheorem{lemma}[thm]{Lemma}
\newtheorem{remark}[thm]{Remark}
\newtheorem{proposition}[thm]{Proposition}

\endlocaldefs
\allowdisplaybreaks
\begin{document}

\newcommand{\boldE}{\mathbf{E}}
\newcommand{\boldP}{\mathbf{P}}
\newcommand{\red}{\textcolor{red}}
\newcommand{\by}{\mathbf{y}}

\setlist[description]{font=\normalfont\space}

\begin{frontmatter}
\title{Asymptotic Distribution-Free Change-Point Detection for Multivariate and Non-Euclidean Data }
\runtitle{Asymptotic Distribution-Free Change-Point Detection} 

\begin{aug}
  \author{\fnms{Lynna}  \snm{Chu} \ead[label=e1]{lbchu@ucdavis.edu}}
  \and
  \author{\fnms{Hao} \snm{Chen}\ead[label=e2]{hxchen@ucdavis.edu}}


  \runauthor{Lynna Chu and Hao Chen}

  \affiliation{University of California, Davis}

  \address{Department of Statistics \\
  University of California, Davis \\
   One Shields Avenue \\
Davis, CA 95616 \\
USA \\
          \printead{e1,e2}}

\end{aug}

\begin{abstract}
We consider the testing and estimation of change-points, locations where the distribution abruptly changes, in a sequence of multivariate or non-Euclidean observations.  We study a nonparametric framework that utilizes similarity information among observations, which can be applied to various data types as long as an informative similarity measure on the sample space can be defined.  The existing approach along this line has low power and/or biased estimates for change-points under some common scenarios.  We address these problems by considering new tests based on similarity information.  Simulation studies show that the new approaches exhibit substantial improvements in detecting and estimating change-points. In addition, under some mild conditions, the new test statistics are asymptotically distribution free under the null hypothesis of no change. Analytic $p$-value approximations to the significance of the new test statistics for the single change-point alternative and changed interval alternative are derived, making the new approaches easy off-the-shelf tools for large datasets. The new approaches are illustrated in an analysis of New York taxi data. 
\end{abstract}



\end{frontmatter}

\section{Introduction}
Change-point analysis is regaining attention as we enter the big data era. Massive amounts of data are collected in many fields for studying complex phenomena over time and/or space. Such data often involve sequences of high-dimensional or non-Euclidean measurements that cannot be analyzed through traditional approaches. Insights on such data often come from segmentation, which divides the sequence into homogeneous temporal or spatial segments. In this paper, we consider this segmentation problem. Let the sequence of observations be $\{ \mathbf{y}_i: i = 1, \hdots, n \}$, indexed by time or some other meaningful orderings. We are concerned with testing the null hypothesis: 
\begin{equation} \label{eq:H0}
 H_0: \, \mathbf{y}_i \sim F_0, \, i = 1, \hdots, n \end{equation} against the single change-point alternative

\begin{equation} \label{eq:H1} 
H_1: \exists \, 1 \le \tau < n, \, \, \mathbf{y}_i \sim 
\begin{cases}
F_0,  \,\,\,\,  i \le \tau \\
F_1,  \,\,\,\, \text{otherwise}
\end{cases}
\end{equation}
or the changed interval alternative 
\begin{equation} \label{eq:H2}
H_2: \exists \, 1 \le \tau_1 < \tau_2 < n, \, \, \mathbf{y}_i \sim 
\begin{cases}
F_0, \,\,\,\, i = \tau_1 + 1, \hdots, \tau_2 \\
F_1, \,\,\,\, \text{otherwise}
\end{cases} \end{equation}
where $F_0$ and $F_1$ are two different probability measures. We consider the problem that observations are independent over time.  (More discussions on violation of this independence assumption can be found in Supplement \ref{sec:bp}.)
 
The segmentation problem has been widely studied for \emph{univariate} data. See monograph \cite{carlstein1994change} for a survey.  However, in many modern applications, $\{\mathbf{y}_i\}$'s could be a sequence of vectors (e.g. cross-sample copy number variation analysis, \cite{zhang2010detecting}), images (e.g. brain image, \cite{park2015anomaly}), or networks (e.g. social network, \cite{Kossinets88}). 

When $\mathbf{y}_i \in \mathbb{R}^d$ and the $d$ dimensions are independent, the problem becomes the analysis of $d$ independent sequences and it has been studied in a number of works, see for examples \cite{zhang2010detecting} and \cite{xie2013sequential}. 
For more generic multivariate observations, most existing methods are based on parametric models (see for examples \cite{chen2011parametric}, \cite{csorgo1997limit} and references therein). 
Parametric methods have been proposed for network data sequences as well. For example, \cite{heard2010bayesian} designed a two-stage Bayesian method to detect anomalies by modeling the communication between nodes over time as a counting process where increments of the process follow a Bayesian probability model. \cite{wang2014locality} designed locality-based scan statistics to detect change arising in the connectivity matrix of networks generated by a stochastic block model where the block membership of the vertices are fixed across time. All these parametric methods provide useful tools when the assumptions made in the paper are reasonably true. However these assumptions are many times too strong in real applications. 

Nonparametric methods have been proposed for the change-point detection problem for multivariate/non-Euclidean observations as well (\cite{jirak2015uniform}, \cite{matteson2014nonparametric}, \cite{lung2011homogeneity}, \cite{cule2010maximum}, \cite{desobry2005online}). Nonparametric methods are usually more flexible in terms of model specification. However, it is in general more difficult to conduct theoretical analysis, such as controlling the type I error. 
 
Recently, \cite{chen2015graph} proposed a non-parametric approach that can be applied to data in arbitrary dimension and to non-Euclidean data. They also provided \emph{analytical p-value approximations} for type I error control, making their approach easy to be applied to large data sets.  Through simulation studies, they showed that their approach achieves substantial power gains when dimension is moderate to high compared with existing parametric change-point methods.  

However, while the method proposed by \cite{chen2015graph} is effective for locational alternatives, it is less effective for scale alternatives and even worse provides biased estimates for the location of the change-point when detected. Also, if the change-point is not in the middle of the sequence, the detection power could be low (more details of these problems are discussed in Section \ref{sec:2}). 

In this paper, we improve upon the limitations of the test statistic in \cite{chen2015graph} and propose three new test statistics. The new test statistics exhibit better estimates to the location of the change-points for a wider range of alternatives and also exhibit substantial power gains when the change is not in the middle of the sequence. In addition, under some mild regularity conditions, the new statistics are asymptotically distribution free under the null hypothesis of no change. The new approaches are implemented in an \texttt{R} package \texttt{gSeg}.

The organization of the rest of the paper is as follows. In Section \ref{sec:2} we describe and explain in more details the problems of the method in  \cite{chen2015graph}. To tackle the problems, three new scan statistics are proposed in Section \ref{sec:3}. The asymptotic behaviors of the new test statistics are studied and analytical $p$-value approximations for the tests are provided in Section \ref{sec:4}. Section \ref{sec:5}  examines the performance of the new test statistics under more simulation settings. The new methods are illustrated in the analysis of New York taxi data in Section \ref{sec:6}.  We conclude with discussion in Section \ref{sec:8}.


\section{Restrictions of the method in Chen and Zhang (2015)}
\label{sec:2}

In this section, we state the restrictions of the method in \cite{chen2015graph} and explore the underlying reasons for these restrictions. 


\subsection{Scenarios when the method breaks down}
\label{sec:2.1}

\begin{figure}[!htp]
\centering
	\includegraphics[height=1\textwidth]{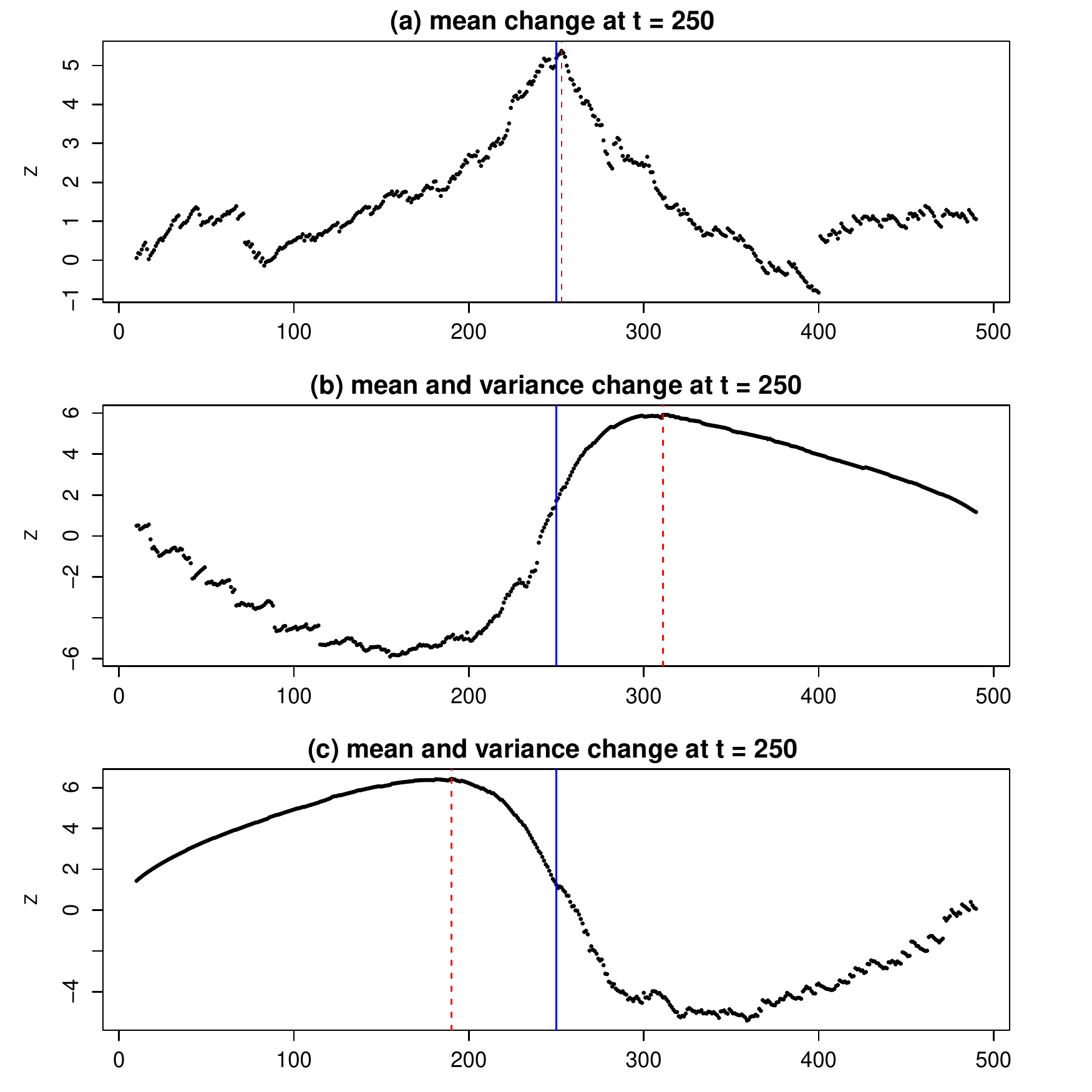}
  \caption{Plots of the scan statistic for the method in \cite{chen2015graph}. Multivariate Gaussian data, $d=100$, $n=500$. Before the change, the data is drawn from $\mathcal{N}(\mathbf{0},\mathbf{I_d})$ and after the change, data drawn from (a) $\mathcal{N}(\mathbf{\mu},\mathbf{I_d})$ where $\|\mu\|_2 = 1.4$, (b) $\mathcal{N}(\mathbf{\mu},\sigma^2 \mathbf{I_d})$ where $\|\mu\|_2 = 1.4$, $\sigma = 1.2$, and (c) $\mathcal{N}(\mathbf{\mu},\mathbf{I_d})$ where $\|\mu\|_2 = 1.4$, $\sigma = 0.8$. The change occurs at $t=250$ for all scenarios. The solid vertical line indicates the true change-point. The dashed vertical line indicates the estimated change-point by the method in \cite{chen2015graph}. The similarity graphs are the $5$-MST constructed using the Euclidean distance. }
  \label{fig:edge_count}
\end{figure}

The method in \cite{chen2015graph} for detecting change-point is a typical scan statistic $\max_{t} Z(t)$, with $Z(t)$ a standardized two-sample test statistic for comparing $\{\mathbf{y}_1, \hdots, \mathbf{y}_t \}$ and $\{\mathbf{y}_{t+1}, \hdots, \mathbf{y}_n \}$. Ideally, when the method works, $Z(t)$ would be maximized around the true change-point. Figure \ref{fig:edge_count} plots $Z(t)$ from typical simulation runs under three different scenarios: (a) a mean change, (b) a change in both mean and variance with the variance larger after the change, and (c) a change in both mean and variance with the variance  smaller after the change.  
 In each scenario, the change occurs at the center of the sequence, indicated by a blue dashed vertical line in each plot.  The estimated change-point is indicated by a black solid vertical line in each plot.   From the plots, the method works perfectly well in scenario (a).  However, it has serious problems in correctly estimating the location of the change-point in scenarios (b) and (c).  From the plots, we see that the estimated change-point is biased towards the direction with a larger variance.


In addition, even when the change is only in mean, the method also has biased change-point estimates along with power loss when the change is not near the middle of the sequence. Table \ref{table:power_example} shows the performance of the method in \cite{chen2015graph} under two choices of the location of the change-point (middle versus one third of the sequence).  It lists the number of trials, out of 100, that the null hypothesis of homogeneity is rejected at the $0.05$ level with the number in the parentheses those trials both rejecting the null and estimating the location of the change-point reasonably well (within 20 indices from the true change-point).   In both settings, the change happens at $\tau=250$ and the change is in the mean only ($\mathcal{N}(\mathbf{0},\mathbf{I_d})$ versus $\mathcal{N}(\mathbf{\mu},\mathbf{I_d})$ where $\|\mu\|_2 = 1.4$ and $d=100$).  

\begin{table}[!htp] 
\caption{The number of trials, out of 100, that the null hypothesis is rejected at 0.05 significance level with the number in the parentheses the number of trials that the null hypothesis is rejected and the index difference between the estimated change-point and true change-point less than 20.   The change happens at $\tau=250$.  The length of the sequence is $n$.  Before the change, the observations are drawn from $\mathcal{N}(\mathbf{0},\mathbf{I_d}), d=100$; after the change, the observations are drawn from $\mathcal{N}(\mathbf{\mu},\mathbf{I_d}), \|\mu\|_2=1.4$.  }

\begin{center}
\begin{tabular}{cc}
\hline
\hline
         $n = 500$  & $n = 750$ \\
\hline
\hline
  97 & 83 \\
       (89) & (24) \\
\hline
\hline
\end{tabular}
\end{center}
\label{table:power_example}
\end{table}

When the length of the sequence is $n=500$, the change happens at the middle of the sequence, and the method does very well.  When $n=750$, since the change happens at $\tau=250$, there are twice as many observations after the change compared to $n=500$.  Intuitively, the increase in sample size should increase the power of the test.  However, different from what we would expect, the performance of the test becomes worse (97 $\rightarrow$ 83).  Even worse is the dramatic decrease in the number in the parentheses ($89\rightarrow$ 24), indicating the poor ability of the the method in estimating the location  of the change-point correctly when the change does not happen in the middle of the sequence.

\subsection{Understanding the graph-based approach}
\label{sec:2.2}
Here, we look closer at the method in \cite{chen2015graph}. It is a scan statistic $Z(t)$ calculated based on a graph-based two-sample test. First, a similarity graph $G$ is constructed on the observations based on a distance measure defined on the sample space up to a criterion. For example, $G$ could be a minimum spanning tree (MST), which is a tree connecting all observations such that the sum of the distances of edges in the tree is minimized; $G$ could also be a nearest neighbor graph (NNG) where each observation connects to its nearest neighbors. Then the number of edges in $G$ that connect observations before $t$ and observations after $t$ are counted. A relative low count indicates the observations before and after $t$ are less mixed, which implies distributional difference. This graph-based two-sample test was first proposed by \cite{friedman1979multivariate} and the intuition behind this is that if the distributions of the two samples are different, observations would tend to be closer to those from the same sample. Thus, edges in the similarity graph would be more likely to connect observations within the same sample. \cite{chen2015graph} adapts this graph-based two-sample test to the change-point setting and $Z(t)$ is a standardized version of the raw count by the mean and standard deviation of the raw count (with a sign flip so that large $Z(t)$ values imply change-points). We refer to this underlying graph-based two-sample test as the \textbf{edge-count two-sample test} for easy reference. 

The rationale of the edge-count two-sample test holds for low-dimensional data. However, when the dimension is high, the edge-count two-sample test can be powerless for some very common types of alternatives due to the curse-of-dimensionality (\cite{chen2017new}). For example, if two distributions differ in variance and when the dimension is moderate to high, such as $d=50$, the two samples would be separated into two layers with the sample with a smaller variance in the inner layer and the other sample in the outer layer. Since the volume of a $d$-dimensional space increases exponentially in $d$, the phenomenon that points in the outer layer find themselves to be closer to points in the inner layer than other points in the outer layer is common unless the number of points in the outer layer is extremely large (exponential in $d$). Then, for typical sample sizes, the between-sample edge-count is still high under this alternative and the edge-count two-sample test is unable to reject the null hypothesis.  To address this issue, \cite{chen2017new} proposed a \textbf{generalized edge-count two-sample test}.

Meanwhile, \citet*{chen2017weighted} found that, starting from the equal sample size scenario, the estimated power of the edge-count two-sample test decreased when one sample size was doubled and the other kept the same. As seen in Table \ref{table:power_example}, even for locational alternatives, this is counter-intuitive since increasing the sample size adds more information, which should increase the power of the test. They found that the decrease in power is due to a variance boosting problem when the sample sizes are unequal.  To address this issue, \citet*{chen2017weighted} proposed a \textbf{weighted edge-count two-sample test}.

In the following, we adapt these two extended graph-based two-sample tests, \textbf{generalized edge-count two-sample test} and the \textbf{weighted edge-count two-sample test}, as well as a new version of the edge-count two-sample test, which we refer to as the \textbf{max-type edge-count two-sample test}, to the change-point setting. 

\section{New test statistics}
\label{sec:3}
The new test statistics for testing the null $H_0$ (\ref{eq:H0}) versus the single change-point alternative $H_1$ (\ref{eq:H1}) and versus the changed interval alternative $H_2$ (\ref{eq:H2}) are presented below. Under the null hypothesis $H_0$ (\ref{eq:H0}), the joint distribution of the observations in the sequence is the same if we permute the order of the observations. In the following, we work under the permutation null distribution that places $1/n!$ probability on each of the $n!$ permutations of $\{ \mathbf{y}_i: i = 1 \hdots n \}$. With no further specification, we use $\boldP, \boldE,$\textbf{Var}, and \textbf{Cov} to denote probability, expectation, variance, and covariance, respectively, under the permutation null distribution. 

\subsection{Generalized edge-count scan statistic for single change-point alternative}
\label{sec:3.1}
Here, we define the test statistic for the generalized edge-count two-sample test when testing the null $H_0$ (\ref{eq:H0}) versus the single change-point alternative $H_1$ (\ref{eq:H1}). 

Each possible value of $\tau$ divides the sequence of observations into two groups: Observations that come before or at $\tau$ and observations that come after $\tau$. Let $G$ be the similarity graph on ${\bf y}_i.$ We use $G$ to denote both the graph and its set of edges when its vertex set is implicitly obvious. For more discussions on the choice of $G$, see \cite{chen2015graph}. For any event $x$ let $I_x$ be the indicator function that takes value $1$ if $x$ is true and $0$ otherwise. We define $g_i(t)$ as an indicator function for the event that $\mathbf{y}_i$ is observed after $t$, $g_i(t) = I_{i > t}$. For an edge $e = (i,j)$, we define 
\begin{align*}
J_e(t) = \begin{cases}
0 \hspace{5mm} \text{ if } g_i(t) \neq g_j(t), \\
1 \hspace{5mm}\text{ if } g_i(t) = g_j(t) = 0, \\
2 \hspace{5mm}\text{ if } g_i(t) = g_j(t) = 1. \\
\end{cases}
\end{align*}
For any candidate value $t$ of $\tau$, we define 
\begin{equation}
R_k(t) = \sum_{e \, \in G} I_{J_e(t) = k}, \, \, \, k = 0,1,2. 
\end{equation}

Then $R_0(t)$ is the number of edges connecting observations before and after $t$ (which is the test statistic for the edge-count two-sample test), $R_1(t)$ is the number of edges connecting observations prior to $t$, and $R_2(t)$ is the number of edges that connect observations after $t$.

\begin{figure}
  \centering
    \includegraphics[width=1\textwidth]{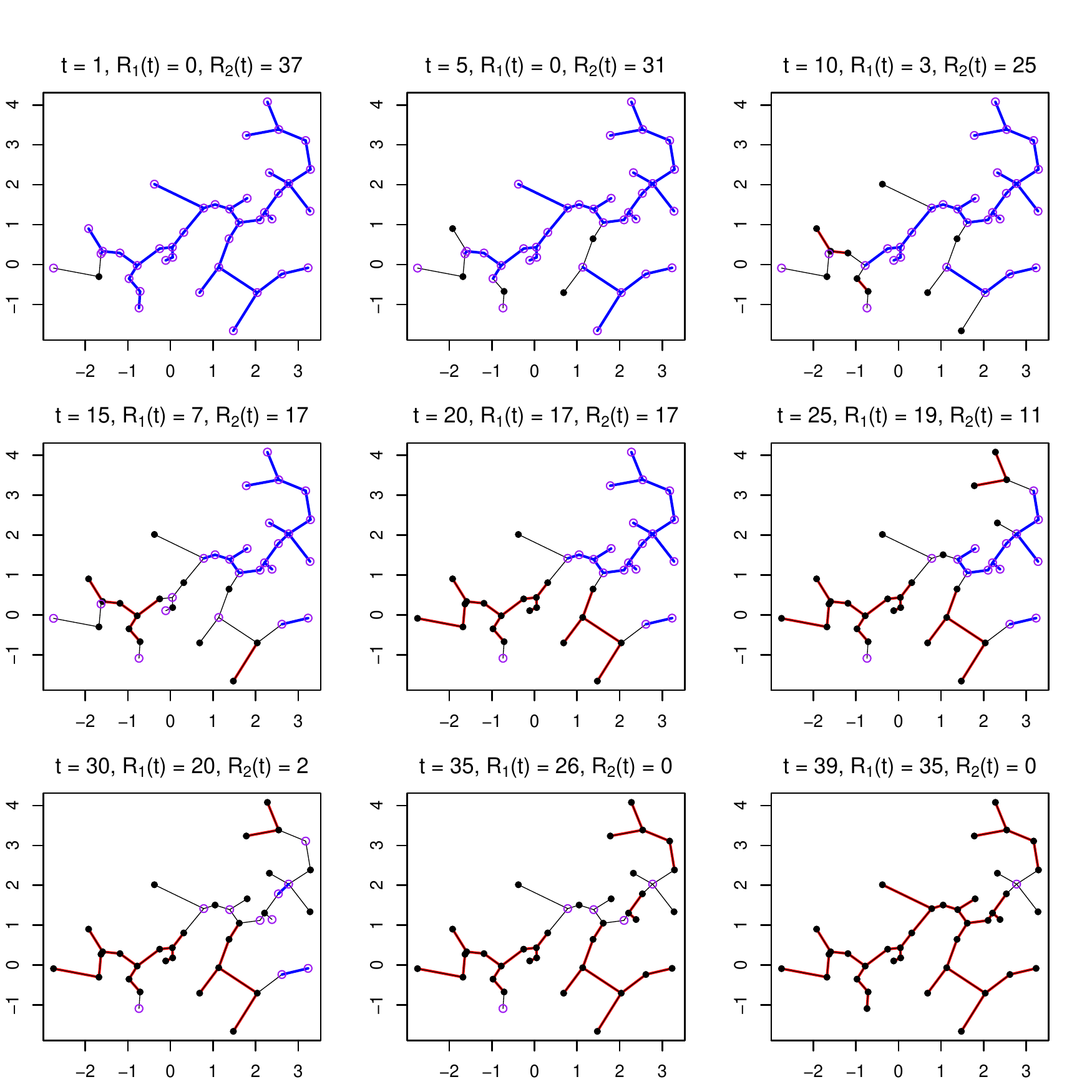}
  \caption{The computation of $R_1(t)$ and $R_2(t)$ for nine different values of $t$. The first 20 observations are generated from $\mathcal{N}(0,I_2)$. The second 20 observations are generated from $\mathcal{N}((2,2)^T,I_2)$. The similarity graph $G$ shown here is the MST on Euclidean distance. Each $t$ divides the observations into two groups: one group for observations before $t$ (shown as solid circles) and the other group for observations shown after $t$ (shown as open circles). Edges in red connect observations before $t$ and the number of these edges is $R_1(t)$. Edges in blue connect observations after $t$ and the number of these edges is $R_2(t)$. Notice that as $t$ changes, the group identities change but the graph $G$ does not change.}
  \label{fig:Rw}
\end{figure}

The generalized edge-count two-sample test at $t$ is defined as 
\begin{equation}
\label{eq:S}
S(t) = \begin{pmatrix}
R_1(t) - \boldE(R_1(t)) \\
R_2(t) - \boldE(R_2(t)) \\
\end{pmatrix}^T \mathbf{\Sigma}^{-1}(t) \begin{pmatrix}
R_1(t) - \boldE(R_1(t)) \\
R_2(t) - \boldE(R_2(t)) \\
\end{pmatrix}. 
\end{equation}
Here, $\mathbf{\Sigma}(t)$ is the covariance matrix of the vector $(R_1(t), R_2(t))^T$ under the permutation null distribution. The test statistic is defined in this way so that either direction of deviations of the number of within-group edges from its null expectation would contribute to the test statistic. Under location alternatives, we would expect both $R_1(t)$ and $R_2(t)$ to be larger than their null expectations, which would lead to a large $S(t)$. Under scale alternatives, the group with the smaller variance would have a within-edge count larger than its null expectation and the group with the larger variance would have a within-edge count smaller than its null expectation, which would also lead to a large $S(t)$. Therefore, this test is powerful for both location and scale alternatives. 

Figure \ref{fig:Rw} illustrates the computation of $R_1(t)$ and $R_2(t)$ on a small artificial dataset of length $n=40$. The first 20 observations are generated from $\mathcal{N}(0,I_2)$. The second 20 observations are generated from $\mathcal{N}((2,2)^T,I_2)$ (the 2-dimensional data is chosen for illustration purposes, while the method is not limited by dimensionality). The similarity graph $G$ is the MST on Euclidean distance. Notice that the graph $G$ is determined by the values of $\mathbf{y}_i$'s and not the order of their appearance. Thus it remains constant under permutation. As $t$ changes, the group identify of some points changes. 

Under the permutation null, the analytic expressions for $\boldE(R_1(t))$, \\ $\boldE(R_2(t))$, and $\mathbf{\Sigma}(t) = (\Sigma_{i,j}(t))_{i,j=1,2}$ can be calculated through combinatorial analysis, and they can be obtained straightforwardly following \cite{chen2017new}. Their expressions are listed below. Let $G_i$ be the subgraph of $G$ containing all edges that connect to node ${\bf y}_i$. Then, $|G_i|$ is the number of edges in $G_i$ or the degree of node ${\bf y}_i$ in $G$. We have
\begin{align*}
\label{eq:summary_S}
&\boldE(R_1(t))  = |G| \tfrac{t (t-1)}{n (n-1)},\\ 
&\boldE(R_2(t))  = |G| \tfrac{(n-t)(n-t-1)}{n (n-1)},\\
&\Sigma_{11}(t)  =  \boldE(R_1(t))(1-\boldE(R_1(t))) +  \tfrac{t(t-1)(t-2)\left(\sum_{i=1}^n |G_i|^2 - 2|G|\right)}{n (n-1)(n-2)}  \\
& \quad \quad \quad \quad + \tfrac{t(t-1)(t-2)(t-3) \left(|G|^2 - \sum_{i=1}^n |G_i|^2  + |G|\right)}{n(n-1)(n-2)(n-3)}, \\
& \Sigma_{22}(t)  =  \boldE(R_2(t))(1-\boldE(R_2(t))) +  \tfrac{(n-t)(n-t-1)(n-t-2)\left(\sum_{i=1}^n |G_i|^2 - 2|G|\right)}{n (n-1)(n-2)} \\
& \quad \quad \quad \quad +  \tfrac{(n-t)(n-t-1)(n-t-2)(n-t-3)\left(|G|^2 - \sum_{i=1}^n |G_i|^2  + |G|\right)}{n (n-1)(n-2)(n-3)},\\
&\Sigma_{12}(t)  = \Sigma_{21}(t) =  \tfrac{t (t-1)(n-t)(n-t-1)\left(|G|^2 - \sum_{i=1}^n |G_i|^2  + |G|\right)}{n (n-1)(n-2)(n-3)} - \boldE(R_1(t))\boldE(R_2(t)).
\end{align*}

To test $H_0$ versus $H_1$, we use the following scan statistic:
\begin{equation} \max_{n_0 \le t \le n_1} S(t), \end{equation}
where $n_0$ and $n_1$ are pre-specified constraints for the range of $\tau$, such as $n_0 = 20$, $n_1 = n - n_0$, as we need some observations in each group to `represent' the distribution. The null hypothesis is rejected if the maxima is greater than a threshold. Details about how to choose the threshold to control the type I error rate are discussed in Section \ref{sec:4}.

\begin{figure}
  \centering
    \includegraphics[width=0.8\textwidth]{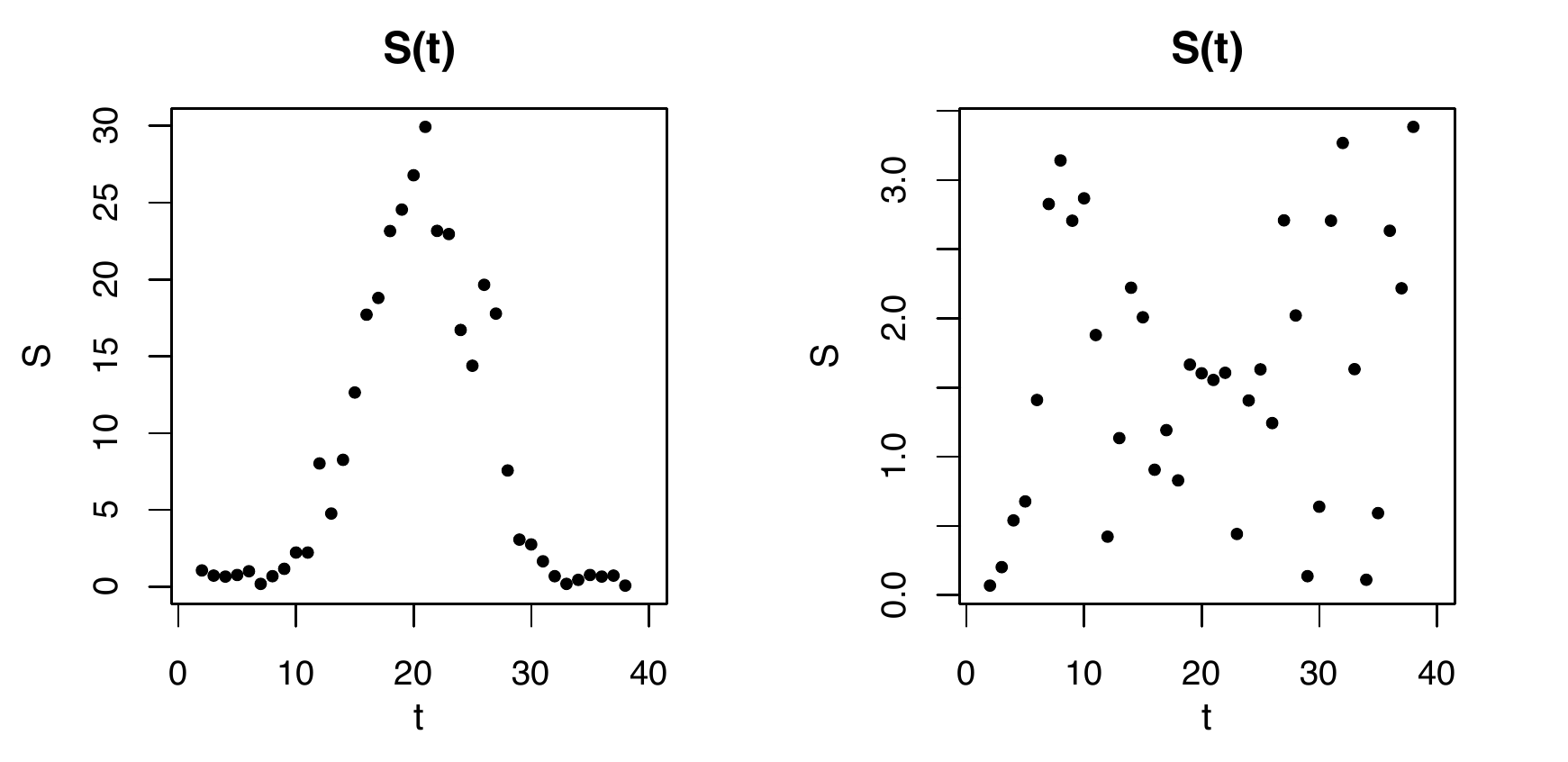}
    \vspace{-2mm}
  \caption{On the left, the profile of $S(t)$ against $t$ for the example data set in Figure \ref{fig:Rw}. On the right, the profile of $S(t)$ against $t$ on a sequence of points with no change-point in a typical simulation run. (The scale on the $y$-axis are different in the two plots.)}
  \label{fig:S_profile}
\end{figure}

Figure \ref{fig:S_profile} shows the $S(t)$ process for the dataset in Figure \ref{fig:Rw} where there is a change-point in the middle (left) and by contrast a typical result when there is no change (right). It is clear that the $\max_{n_0 \le t \le n_1} S(t)$ in the left panel is much larger. 

\subsection{Weighted edge-count scan statistic for single change-point alternative}
\label{sec:3.2}

Here, we present the weighted edge-count two-sample test statistic for testing the null $H_0$ (\ref{eq:H0}) versus the single change-point alternative $H_1$ (\ref{eq:H1}). Following the same notations in Section \ref{sec:3.1}, for any candidate value $t$ of $\tau$, the weighted edge-count two-sample test statistic is 
\begin{align*} 
R_w(t) &  = q(t) \sum_{e \in G} I_{J_e(t)=1} + p(t) \sum_{e \in G} I_{J_e(t)=2}  = q(t) \, R_1(t) + p(t) \, R_2(t), 
\end{align*}
where $p(t) = \frac{t-1}{n-2}$ and $q(t)  = 1 - p(t)$. As it is more difficult for the sample with a smaller sample size to form an edge within the sample, $R_1(t)$ and $R_2(t)$ are weighted by the inverse of their corresponding sample sizes. The test statistic defined in this way resolves the variance boosting problem \citep{chen2017weighted}. Relatively large values of $R_w(t)$ are evidence against the null hypothesis. 

Since the null distribution of $R_w(t)$ depends on $t$, $R_w(t)$ is standardized so that it is comparable across $t$. Let
\begin{equation} 
\label{eq:Zw}
Z_w(t) = \frac{R_w(t) - \boldE[R_w(t)]}{\sqrt{\text{\bf Var}[R_w(t)]}}.
\end{equation}

Analytic formulas for $\boldE(R_w(t))$ and $\text{\bf Var}(R_w(t))$ are given below:
\begin{align*}
\label{eq:summary_Rw}
 \boldE(R_w(t)) & = |G| \tfrac{(t-1)(n-t-1)}{(n-1)(n-2)}. \\
 \text{\bf{Var}}(R_w(t)) & =
\tfrac{t(t-1)(n-t)(n-t-1)}{n (n-1)(n-2)(n-3)}  
 \left ( |G| -  \tfrac{\sum_{i=1}^n |G_i|^2}{(n-2)} + \tfrac{2|G|^2}{(n-1)(n-2)} \right).
\end{align*}
To test $H_0$ versus $H_1$, the following scan statistic is used: 
\begin{equation} \max_{n_0 \le t \le n_1} Z_w(t), \end{equation}
where $n_0$ and $n_1$ are pre-specified constraints for the range of $\tau$. The null hypothesis is rejected if the maxima is greater than a threshold. Details about how to choose the threshold to control the type I error are discussed in Section \ref{sec:4}. 

\begin{figure}[htp]
  \centering
    \includegraphics[width=0.8\textwidth]{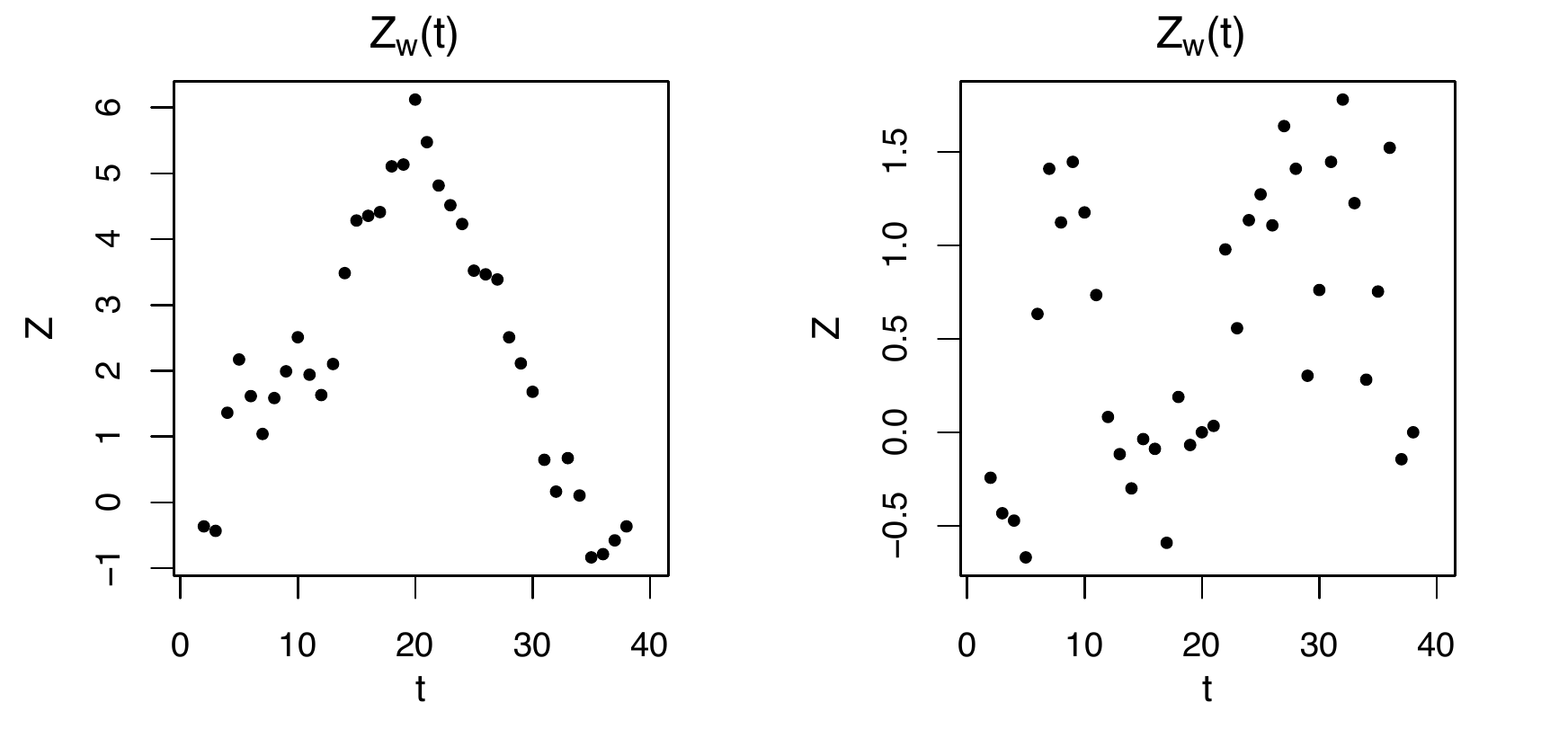}
  \vspace{-3mm}
    \caption{On the left, the profile of $Z_w(t)$ against $t$ for the example data set in Figure \ref{fig:Rw}. On the right, the profile of $Z_w(t)$ against $t$ on a sequence of points with no change-point. (The scale on the $y$-axis are different in the two plots.)}

  \label{fig:Zw_profile}
\end{figure}

For illustration, Figure \ref{fig:Zw_profile} shows the $Z_w(t)$ processes for the same illustration dataset as in Figure \ref{fig:Rw}. We see that $Z_w(t)$ peaks at the true change-point $\tau = 20$. For contrast, when there is no change-point, $Z_w(t)$ exhibits random fluctuation and attains a much smaller maximum value compared to when there is a change-point. 
\subsection{Scan statistics for changed interval alternative}
For testing the changed interval alternative $H_2$ (\ref{eq:H2}), each possible interval $(t_1,t_2]$ partitions the observations into two groups: one group containing all observations observed during $(t_1, t_2]$, and the other group containing all observations observed outside of this interval. Then, for any candidate changed interval $(t_1, t_2]$, we have that $R_0(t_1,t_2)$ is the number of edges connecting observations within and outside the interval $(t_1,t_2]$, $R_1(t_1,t_2)$ is the number of edges connecting observations outside of the interval $(t_1,t_2]$, and $R_2(t_1,t_2)$ is the number of edges connecting observations within the interval $(t_1,t_2]$.  Then the two-sample test statistics for testing the changed interval alternative can be defined in a similar manner to the single change-point case in Sections \ref{sec:3.1} and \ref{sec:3.2}. For example, the generalized edge-count two-sample test statistic, $S(t_1,t_2)$, for testing $H_0$ (\ref{eq:H0}) versus $H_2$ (\ref{eq:H2}) is defined as 
\begin{align*}
  \binom{R_1(t_1,t_2) - \boldE(R_1(t_1,t_2))}{R_2(t_1,t_2) - \boldE(R_2(t_1,t_2))}^T
\mathbf{\Sigma}^{-1}(t_1,t_2) \binom{R_1(t_1,t_2) - \boldE(R_1(t_1,t_2))}{R_2(t_1,t_2) - \boldE(R_2(t_1,t_2))}.
\end{align*}

Under the permutation null, the explicit expression for $\boldE(R_1(t_1,t_2))$, \\ $\boldE(R_2(t_1,t_2))$ and the covariance matrix can be obtained similarly as in the single change-point setting. The explicit expressions can be found in Supplement \ref{app:A}.  The scan statistic involves a maximization over $t_1$ and $t_2$, i.e., 
 \begin{equation} \max_{\substack{\\ 1 \le t_1 < t_2 \le n \\ l_0 \le t_2 - t_1 \le l_1}} S(t_1,t_2), \end{equation}
where $l_0$ and $l_1$ are constraints on the window size. For example, we can set $l_1 = n - l_0$ so that only alternatives where the numbers of observations in either group is larger than $l_0$ are considered. 

Complete details of the generalized edge-count scan statistic and weighted edge-count scan statistic for the changed alternative are given in Supplement \ref{app:A}. 

\subsection{Max-type edge-count two-sample test}
\label{sec:3.5}
Here we present a new test statistic, based on the following lemma: 
\begin{lemma}
\label{thm:S_finite} The generalized edge-count scan statistic can be expressed as 
\[S(t) = Z_w^2(t)+ Z_\text{diff}^2(t),\] 
\[S(t_1,t_2)=Z_w^2(t_1,t_2)+ Z_\text{diff}^2(t_1,t_2),\] where $Z_w(t)$, $Z_w(t_1,t_2)$ are the standardized weighted edge-count two-sample test statistic defined in (\ref{eq:Zw}) and in Supplement \ref{app:A}, respectively, and 
\begin{align} 
\label{eq:Zdiff} Z_\text{diff}(t) & = \frac{R_{\text{diff}}(t)- \boldE(R_{\text{diff}}(t))}{\sqrt{\text{\bf Var}(R_{\text{diff}}(t))} }, \\
\label{eq:Zdiff2} Z_\text{diff}(t_1,t_2) & = \frac{R_{\text{diff}}(t_1,t_2) - \boldE(R_{\text{diff}}(t_1,t_2))}{\sqrt{\text{\bf Var}(R_{\text{diff}}(t_1,t_2))}},
\end{align}
with $R_{\text{diff}}(t) = R_1(t)-R_2(t)$ and $R_{\text{diff}}(t_1,t_2) = R_1(t_1,t_2)-R_2(t_1,t_2)$.
\end{lemma}

The proof of this lemma is in Supplement \ref{app:Proofs}.  The analytical expressions for the expectation and variance of $R_{\text{diff}}(t)$ and $R_{\text{diff}}(t_1,t_2)$ under the permutation null are: 
\begin{align*}
 \boldE(R_\text{diff}(t)) & = |G|\frac{(2t-n)}{n}, \\
 \boldE(R_\text{diff}(t_1,t_2)) & =  |G| \frac{(2(t_2 - t_1)-n)}{n}, \\
  \text{\bf{Var}}(R_\text{diff}(t)) & = \frac{t(n-t)\left(\sum_{i=1}^n |G_i|^2 - \tfrac{4|G|^2}{n} \right)}{n(n-1)}, \\
 \text{\bf{Var}}(R_\text{diff}(t_1,t_2)) & = \frac{(t_2-t_1)(n+t_2-t_1)\left(\sum_{i=1}^n |G_i|^2 - \tfrac{4|G|^2}{n} \right)}{n(n-1)}.
\end{align*}

From the above lemma, we can see that $S(t)$ is the sum of squares of two uncorrelated quantities (these two quantities are further asymptotically independent, see in Section \ref{sec:4}). Here, $Z_w(t)$ tends to be sensitive to locational alternatives. When the change is locational, $Z_w(t)$ tends to be large. On the other hand, $Z_\text{diff}(t)$ tends to be sensitive to scale alternative. When the change is in the spread of the distribution, $|Z_\text{diff}(t)|$ tends to be large. The sign of $Z_\text{diff}(t)$ depends on whether the distribution after the change has a larger spread or not. Hence, we propose the following max-type edge-count two-sample test statistic: 
\begin{equation}
\label{eq:M}
M(t) = \max \left(|Z_\text{diff}(t)|,Z_w(t)\right) 
\end{equation}
for the single change-point alternative and  
\begin{equation}
\label{eq:M2}
M(t_1,t_2) = \max \left(|Z_\text{diff}(t_1,t_2)|,Z_w(t_1,t_2)\right) 
\end{equation}
for the changed-interval alternative. The corresponding scan statistics are
\begin{equation}
\label{eq:scan_M} \max_{n_0 \le t \le n_1} M(t),
\end{equation}
for the single change-point alternative and
\begin{equation} 
\label{eq:scan_M2} \max_{\substack{\\ 1 \le t_1 < t_2 \le n \\ l_0 \le t_2 - t_1 \le l_1}} M(t_1,t_2), \end{equation}
for the changed-interval alternative.  

As it will come later, this max-type statistic is of particular interest as its performance is similar to $S(t)$ and we can obtain more accurate $p$-value approximations (details in Section \ref{sec:4}). 

A more detailed discussion on the relationship of the three test statistics ($S, Z_w, M$) and an extension to the max-type statistic can be found in Supplement \ref{sec:Mk}.

\section{Analytical p-value approximations}
\label{sec:4}
Given the scan statistics, the next question is how large do they need to be to constitute sufficient evidence against the null hypothesis of homogeneity. In order words, we are concerned with the tail probability of the scan statistics under $H_0$. For the generalized edge-count two-sample test, that is, 
\begin{equation} 
\label{eq:tail_S}
\boldP \left(\max_{n_0 \le t \le n_1} S(t) > b \right) \end{equation}
for the single change-point alternative, and 
\begin{equation}
\label{eq:tail_S2}
\boldP \left(\max_{\substack{\\ 1 \le t_1 < t_2 \le n \\ l_0 \le t_2 - t_1 \le l_1}} S(t_1,t_2) > b \right)  \end{equation}
for the changed interval alternative. For the weighted edge-count two-sample test, that is, 
\begin{equation} 
\label{eq:tail_Rw}
\boldP \left(\max_{n_0 \le t \le n_1} Z_w(t) > b \right) 
\end{equation}
for the single change-point alternative, and 
\begin{equation}
\label{eq:tail_Rw2} 
\boldP \left(\max_{\substack{\\ 1 \le t_1 < t_2 \le n \\ l_0 \le t_2 - t_1 \le l_1}} Z_w(t_1,t_2) > b \right)  \end{equation}
for the changed interval alternative. For the max-type edge-count two-sample test, that is, 
\begin{equation} 
\label{eq:tail_M}
\boldP \left(\max_{n_0 \le t \le n_1} M(t) > b \right) 
\end{equation}
for the single change-point alternative, and 
\begin{equation}
\label{eq:tail_M2} 
\boldP \left(\max_{\substack{\\ 1 \le t_1 < t_2 \le n \\ l_0 \le t_2 - t_1 \le l_1}} M(t_1,t_2) > b \right)  \end{equation}
for the changed interval alternative. 

For small $n$, we can directly sample from the permutation distribution to approximate (\ref{eq:tail_S}) - (\ref{eq:tail_M2}). However, when $n$ is large, permutation is very time consuming. Therefore, to make the method instantly applicable, we derive analytical expressions to approximate these tail probabilities. 

To derive the analytical expressions, we study the asymptotic properties of the stochastic processes $\{S(t)\}$, $\{S(t_1,t_2)\}$, $\{Z_w(t)\}$, $\{Z_w(t_1,t_2)\}$, $\{M(t)\}$, and $\{M(t_1,t_2)\}$, and then make adjustments for finite samples. By Lemma \ref{thm:S_finite} and how $M(t)$ is defined, these stochastic processes boil down to two pairs of basic processes: $\{Z_\text{diff}(t)\}$ and $\{Z_w(t)\}$ for the single change-point case and $\{Z_\text{diff}(t_1,t_2)\}$ and $\{Z_w(t_1,t_2)\}$ for changed-interval. So we first study the properties of these basic stochastic processes. 

\subsection{Asymptotic null distributions of the basic processes}
\label{sec:4.1}

In this section, we derive the limiting distributions of $\{Z_\text{diff}([nu]):0 < u < 1\}$ and\\ $\{Z_w([nu]):0 < u < 1\}$ for the single change-point alternative, and \\$\{Z_\text{diff}([nu],[nv]):0 < u < v< 1\}$ and $\{Z_w([nu],[nv]):0 < u < v< 1\}$ for the changed-interval alternative (we use $[x]$ to denote the largest integer that is no larger than $x$).

We first introduce some notations. For edge $e = (e_{-},e_{+})$, where $e_{-} < e_{+}$ are the indices of the nodes connected by the edge $e$, let
\begin{equation} \label{eq:Ae}
A_e = G_{e_{-}} \cup G_{e_{+}},\end{equation}
be the subgraph in $G$ that connect to either node $e_{-}$ or node $e_{+}$, and 
\begin{equation} \label{eq:Be}
B_e = \underset{e^* \in A_e}{\cup} \, A_{e^*} , \end{equation}
be the subgraph in $G$ that connect to any edge in $A_e$. 

In the following, we write $a_n = O(b_n)$ when $a_n$ has the same order as $b_n$, and write $a_n = o(b_n)$ when $a_n$ has order smaller than $b_n$.


\begin{thm}
\label{thm:GausProc}
When  $|G| = O(n^\alpha), 1 \le \alpha < 1.5$, $\sum_{e \in G} |A_e||B_e| = o(n^{1.5 \alpha})$, 
$\sum_{e \in G} |A_e|^2 = o(n^{\alpha+0.5})$, and $\sum_{i=1}^n |G_i|^2 - \tfrac{4|G|^2}{n} = O(\sum_{i=1}^n |G_i|^2)$, as $n \rightarrow \infty$,
\begin{enumerate}
\item $\{ Z_\text{diff}([nu]): 0< u <1\}$ and $\{Z_w([nu]):0<u<1 \}$ converge to independent Gaussian processes in finite dimensional distributions, which we denote as $\{Z_\text{diff}^*(u): 0 < u < 1\} $ and $\{Z_w^*(u):0 < u < 1\}$, respectively. 
\item $\{ Z_\text{diff}([nu],[nv]): 0< u < v< 1\}$ and $\{ Z_w([nu],[nv]): 0< u < v< 1\}$ converge to independent two-dimension Gaussian random fields in finite dimensional distributions, which we denote as $\{Z_\text{diff}^*(u,v): 0 < u < v<  1\}$ and $\{Z_w^*(u,v): 0 < u < v<  1\}, $ respectively. 
\end{enumerate}
\end{thm}

The proof for this theorem utilizes Stein's method (\cite{ChenShao2005}) and the details of the proof are in Supplement \ref{app:Proofs}.

\begin{remark}
The condition $|G| = O(n^\alpha), 1 \le \alpha < 1.5$ ensures that the graph is dense enough but not too dense. The conditions $\sum_{e \in G} |A_e||B_e| = o(n^{1.5 \alpha})$ and $\sum_{e \in G} |A_e|^2 = o(n^{\alpha+0.5})$ ensure that the graph does not have a large hub or a cluster of small hubs, where a hub is a node with a large degree. The condition $\sum_{i=1}^n |G_i|^2 - \tfrac{4|G|^2}{n} = O(\sum_{i=1}^n |G_i|^2)$ ensures $Z_\text{diff}$ to be well-defined. 

These conditions are quite mild. For example, 
for $k$-MST, when $k=O(1)$, we have $|G|=k(n-1) = O(n)$, and the conditions boil down to $\sum_{e \in G} |A_e||B_e| = o(n^{1.5})$ and $\sum_{i=1}^n |G_i|^2 - \tfrac{4|G|^2}{n} = O(\sum_{i=1}^n |G_i|^2)$.  Based on Theorems 5.1 and 5.2 in \cite{chen2017new}, both conditions are satisfied for $k$-MST constructed on Euclidean distance for $k=O(1)$.  

More discussions on the conditions of the graph can be found in Supplement \ref{sec:condition}.

\end{remark}


Let $\rho^*_w(u,v) = \textbf{Cov}(Z^*_w(u),Z^*_w(v))$ and $\rho^*_\text{diff}(u,v) = \textbf{Cov}(Z^*_\text{diff}(u),Z^*_\text{diff}(v))$.  The next theorem state explicitly the covariance functions of the limiting Gaussian processes, $\{Z^*_w(u), \, 0 < u < 1 \}$ and $\{Z^*_\text{diff}(u), \, 0 < u < 1 \}$.

\begin{thm}
\label{thm:cov_S}
The exact expressions for $\rho^*_\text{diff}(u,v)$ and $\rho^*_w(u,v)$ are: 
\begin{align*}
\rho^*_w(u,v) & = \frac{(u \wedge v)(1-(u \vee v))}{(u \vee v)(1 - (u \wedge v))}, \\
\rho^*_\text{diff}(u,v) & = \frac{(u \wedge v)(1- (u \vee v))}{\sqrt{(u \wedge v) (1 - (u \wedge v))(u \vee v)(1-(u \vee v))}},
\end{align*}
where $ u\wedge v = \min(u,v)$ and $u\vee v = \max(u,v)$. 
\end{thm}

The above theorem is proved through combinatorial analysis and details are given in the Supplement \ref{app:Proofs} . From the above theorem, we see that the limiting processes, $\{Z^*_w(u), \, 0 < u < 1 \}$ and $\{Z^*_\text{diff}(u), \, 0 < u < 1 \}$, do not depend on $G$ at all.


\subsection{Asymptotic $p$-value approximations}
We now examine the asymptotic behavior of the tail probabilities (\ref{eq:tail_S}) - (\ref{eq:tail_M2}). Our approximations require the function $\nu(x)$ defined as
\begin{equation}
\label{eq:nu} \nu(x) = 2x^{-2} \exp \left ( -2 \sum_{m=1}^{\infty}m^{-1}\Phi \left( -\tfrac{1}{2}xm^{1/2} \right)\right ), x > 0.\end{equation}
This function is closely related to the Laplace transform of the overshoot over the boundary of a random walk. A simple approximation given in \cite{siegmund2007statistics} is sufficient for numerical purpose: 
\begin{equation}
\label{eq:nu_approx} \nu(x) \approx \frac{(2/x)(\Phi(x/2)-0.5)}{(x/2)\Phi(x/2)+\phi(x/2)},\end{equation}
where $\Phi(\cdot)$ and $\phi(\cdot)$ denote the standard normal cumulative density function and standard normal density function, respectively. Following similar arguments in the proof for Proposition 3.4 in \cite{chen2015graph}, when the conditions on $G$ in Theorem \ref{thm:GausProc} hold, $n,b,n_0,n_1 \rightarrow \infty$ in a way such that for some $b_0>0$ and $0<x_0<x_1<1$, $b/\sqrt{n}\rightarrow b_0$, $\tfrac{n_0}{n}\rightarrow x_0$ and $\tfrac{n_1}{n}\rightarrow x_1$, 
then as $n \rightarrow \infty$, we have 
\begin{align*}
& \boldP\left(\max_{n_0 \le t \le n_1} Z^*_w(t/n) > b\right) \sim \, b \phi(b) \int_{x_0}^{x_1} h_w^*(x) \nu (b_0 \sqrt{2h^*(x)})dx, \\
& \boldP \left(\max_{{n_0 \le t_2 -t_1 \le n_1}} Z^*_w({t_1}/{n},{t_2}/{n}) > b \right)  \\
& \hspace{30mm}   \sim b^3 \phi(b) \int_{x_0}^{x_1} \left( h_w^*(x) \nu (b_0 \sqrt{2h_w^*(x)}) \right ) ^2 (1-x)dx \nonumber \\
& \boldP \left(\max_{n_0 \le t \le n_1} |Z_\text{diff}^*(t/n)| > b \right)   \sim 2 b \phi(b) \int_{x_0}^{x_1} h^*_\text{diff}(x) \nu (b_0 \sqrt{2 h^*_\text{diff}(x)}) dx, \nonumber \\
& \boldP \left(\max_{n_0 \le t_2 -t_1 \le n_1} |Z^*_\text{diff}(t_1/n,t_2/n)| > b \right)  \\
& \hspace{30mm}   \sim 2b^3 \phi(b) \int_{x_0}^{x_1} \left( h_\text{diff}^*(x) \nu (b_0 \sqrt{2h_\text{diff}^*(x)}) \right ) ^2 (1-x)dx, \nonumber
\end{align*}
where 
\begin{align*}
 h^*_w(x) & = \lim_{u\nearrow x} \frac{\partial \rho_w^*(u,x)}{\partial u} \equiv -\lim_{u\searrow x} \frac{\partial \rho_w^*(u,x)}{\partial u}, \\
 h^*_\text{diff}(x) & = \lim_{u\nearrow x} \frac{\partial \rho_\text{diff}^*(u,x)}{\partial u} \equiv -\lim_{u\searrow x} \frac{\partial \rho_\text{diff}^*(u,x)}{\partial u} .
\end{align*}
It can be shown that
\begin{align}
h^*_w(x) & = \frac{1}{x(1-x)}, \label{eq:hw} \\
h^*_{\text{diff}}(x) & = \frac{1}{2x(1-x)}. \label{eq:hdiff}
\end{align}

Since $Z_w^*$ and $Z_{\text{diff}}^*$ are independent, we have 
\begin{align*}
\boldP & \left(\max_{n_0 \le t \le n_1} M^*(t/n) > b \right) \\
& =1 - \boldP \left( \max_{n_0 \le t \le n_1} |Z^*_\text{diff}(t)| < b \right) \boldP \left( \max_{n_0 \le t \le n_1} Z^*_w(t) < b \right), \nonumber \\
\boldP & \left( \underset{n_0 \le t_2 -t_1 \le n_1}{\max} M^*(t_1/n,t_2/n) > b \right)  \\
 & =1 - \boldP \left( \max_{n_0 \le t_2 -t_1 \le n_1} |Z^*_\text{diff} (t_1,t_2)| < b \right) \boldP \left( \max_{n_0 \le t_2 -t_1 \le n_1} Z^*_w(t_1,t_2) < b \right). \nonumber
\end{align*}


For the tail probabilities for $\max_{n_0 \le t \le n_1} S(t)$ and $\max_{l_0 \le t_2 - t_2 \le l_1} S(t_1,t_2)$, some additional works are needed and the results are stated in the following proposition.   
\begin{proposition}
\label{thm:S_pvalue_approx} Assume that  $|G| = O(n^\alpha), 1 \le \alpha < 1.5$, $\sum_{e \in G} |A_e||B_e| = o(n^{1.5 \alpha})$, 
$\sum_{e \in G} |A_e|^2 = o(n^{\alpha+0.5})$, and $\sum_{i=1}^n |G_i|^2 - \tfrac{4|G|^2}{n} = O(\sum_{i=1}^n |G_i|^2)$, $n,b,n_0,n_1 \rightarrow \infty$ in a way such that for some $b_1>0$ and $0<x_0<x_1<1$, $b/n\rightarrow b_1$, $\tfrac{n_0}{n}\rightarrow x_0$ and $\tfrac{n_1}{n}\rightarrow x_1$, 
then as $n \rightarrow \infty$,
\begin{align}
\boldP & \left(\max_{n_0 \le t \le n_1} S^*(t/n) > b \right) \\
& \hspace{5mm}  \approx \frac{b\,e^{-b/2}}{2\pi}  \int_{0}^{2 \pi} \int_{x_0}^{x_1} u^*(x,\omega) \nu (\sqrt{2 b_1 \, u^*(x,\omega)}) dx d \omega \nonumber \\
\boldP & \left (\max_{n_0 \le t_2 - t_1 \le n_1} S^*(t_1/n, t_2/n) > b \right)   \\
& \hspace{5mm} \approx \frac{b^2 e^{-b/2}}{\pi}  \int_{0}^{2 \pi} \int_{x_0}^{x_1} \left( u^*(x,\omega)   \nu(\sqrt{2b_1 \, u^*(x,\omega)}\, \, ) \right)^2 (1-x) dx d\omega   \nonumber
\end{align}
where 
$ u^*(x,\omega) =  h_w^*(x)\sin^2(\omega) + h_\text{diff}^*(x)\cos^2(\omega),$
with $h^*_w(x)$ and $h^*_{\text{diff}}(x)$ provided in \eqref{eq:hw} and \eqref{eq:hdiff}, respectively.
\end{proposition}
The proof of this proposition is in Supplement \ref{app:Proofs}. 

Based on the above results, we can approximate the tail probabilities (\ref{eq:tail_S}) - (\ref{eq:tail_M2}) by
\begin{align}
& \boldP \left(\max_{n_0 \le t \le n_1} S(t) > b \right)  \label{eq:S_approx} \\
& \hspace{1mm}  \approx \frac{b\, e^{-b/2}}{2\pi} \int_{0}^{2 \pi} \int_{\tfrac{n_0}{n}}^{\tfrac{n_1}{n}}    u^*(x,\omega) \nu (\sqrt{2 b \, u^*(x,\omega)/n}) dx d \omega, \nonumber \\
& \boldP \left (\max_{n_0 \le t_2 - t_1 \le n_1} S(t_1, t_2) > b \right)  \label{eq:S2_approx} \\
&  \hspace{1mm} \approx \frac{b^2 e^{-b/2}}{\pi} \int_{0}^{2 \pi} \int_{\tfrac{n_0}{n}}^{\tfrac{n_1}{n}} \left( u^*(x,\omega)   \nu(\sqrt{2b \, u^*(x,\omega)/n}\, \, ) \right)^2 (1-x) dx d\omega, \nonumber \\
& \boldP\left(\max_{n_0 \le t \le n_1} Z_w(t) > b\right) \approx \, b \phi(b) \int_{\tfrac{n_0}{n}}^{\tfrac{n_1}{n}} h_w^*(x) \nu (b \sqrt{2h^*(x)/n})dx \label{eq:Zw_approx} \\
& \boldP \left(\max_{n_0 \le t_2 -t_1 \le n_1} Z_w(t_1,t_2) > b \right) \label{eq:Zw2_approx} \\
& \hspace{1mm}  \approx b^3 \phi(b) \int_{\tfrac{n_0}{n}}^{\tfrac{n_1}{n}} \left( h_w^*(x) \nu (b \sqrt{2h_w^*(x)/n}) \right ) ^2 (1-x)dx, \nonumber \\
& \boldP \left(\max_{n_0 \le t \le n_1} M(t) > b \right) \label{eq:M_approx}\\
& \hspace{1mm} = 1 - \boldP \left( \max_{n_0 \le t \le n_1} |Z_\text{diff}(t)| < b \right) \boldP \left( \max_{n_0 \le t \le n_1} Z_w(t) < b \right), \nonumber\\ 
& \boldP \left( \max_{n_0 \le t_2 -t_1 \le n_1} M(t_1,t_2) > b \right)   \label{eq:M2_approx}\\
\hspace{5mm} = 1 - &  \boldP \left( \max_{n_0 \le t_2 -t_1 \le n_1} |Z_\text{diff} (t_1,t_2)| < b \right) \boldP \left( \max_{n_0 \le t_2 -t_1 \le n_1} Z_w(t_1,t_2) < b \right), \nonumber
\end{align}
where,
\begin{align}
& \boldP \left(\max_{n_0 \le t \le n_1} |Z_\text{diff}(t)| < b \right) \label{eq:Zdiff_approx} \\
&\hspace{14mm} \approx 1-2b \phi(b) \int_{\tfrac{n_0}{n}}^{\tfrac{n_1}{n}} h^*_\text{diff}(x) \nu (b \sqrt{2 h^*_\text{diff}(x)/n}) dx, \nonumber \\
& \boldP \left(\max_{n_0 \le t_2 -t_1 \le n_1} |Z_\text{diff}(t_1,t_2)| < b \right)  \label{eq:Zdiff2_approx} \\
& \hspace{14mm} \approx 1- 2b^3 \phi(b) \int_{\tfrac{n_0}{n}}^{\tfrac{n_1}{n}} \left( h_\text{diff}^*(x) \nu (b \sqrt{2h_\text{diff}^*(x)/n}) \right ) ^2 (1-x)dx, \nonumber 
\end{align}
and $\boldP \left(\max_{n_0 \le t \le n_1} Z_w(t) < b \right)$ and  $\boldP \left(\max_{n_0 \le t_2 - t_1 \le n_1} Z_w(t_1, t_2) < b \right)$ easily follow from (\ref{eq:Zw_approx}) and (\ref{eq:Zw2_approx}), respectively. 

\begin{remark}
In practice, when using (\ref{eq:S_approx}) - (\ref{eq:M2_approx}) to approximate the tail probabilities,
we use $h_w(n, x)$ in place of $h^*_w(x)$, where $h_w(n, x)$ is the finite-sample equivalent of $h^*_w(x)$. That is, 
\begin{align*}
h_w(n,x) = n\lim_{s\nearrow nx} \frac{\partial \rho_w(s,nx)}{\partial s},
\end{align*}
with $\rho_w(s,t):= \textbf{Cov} (Z_w(s),Z_w(t))$.
The explicit expression for $h_w(n,x)$ can also be derived and simplified to be
\begin{equation}
\label{eq:hwn}
h_w(n,x)= \frac{(n - 1)(2n x^2 - 2nx + 1)}{2x (1-x)(n^2x^2 - n^2x + n - 1)}.\\
\end{equation}
It is clear from the above expression that $h_w(n,x)$ does not depend on the graph $G$ as well.  Also, it is easy to show that $\lim_{n \rightarrow \infty} h_w(n,x) = h^*_w(x)$.

The finite-sample equivalent version of $h^*_\text{diff}(x)$ is exact the same as $h^*_\text{diff}(x)$.  That is,
\begin{align*}
h_\text{diff}(n,x) & = n \lim_{s\nearrow nx} \frac{\partial \textbf{Cov} (Z_\text{diff}(s),Z_\text{diff}([nx]))}{\partial s} = \frac{1}{2x(1-x)}.
\end{align*}   

%
%
\end{remark}

\subsection{Skewness Correction}

\begin{figure}[htp]
  \centering
	    \includegraphics[width=0.8\textwidth]{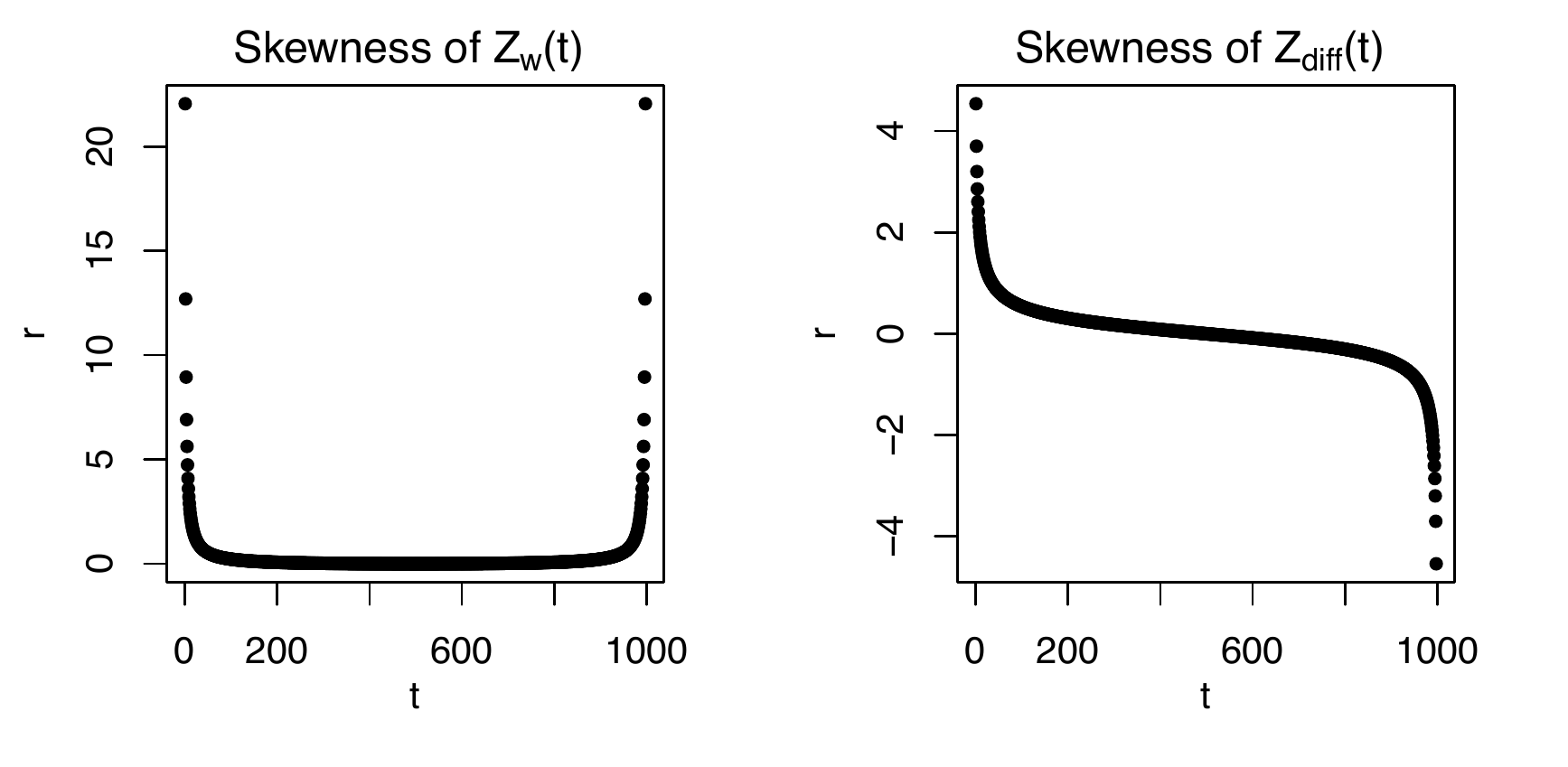}
      \vspace{-2mm}
	    \caption{\small Plots of skewness of $Z_w(t)$ and of $Z_\text{diff}(t)$ against $t$ for a sequence of 1,000 points randomly generated from $\mathcal{N}(0,I_{100})$. The graph is MST constructed on Euclidean distance.}
	    \label{fig:skew}
\end{figure}

Analytical approximations become less precise when the minimum window length decreases (see numerical results in Section \ref{sec:4.4}). This is mainly because the convergence of $Z_w(t)$ and $Z_\text{diff}(t)$ to normal is slow if $t/n$ is close to $0$ or $1$ and the convergence of $Z_w(t_1,t_2)$ and $Z_\text{diff}(t_1,t_2)$ to normal is slow if $\frac{t_2 - t_1}{n}$ is close to $0$ or $1$. 
This problem becomes more severe when dimension is high. 
Figure \ref{fig:skew} plots the skewness of $Z_w(t)$ and $Z_\text{diff}(t)$ with $G$ being MST constructed on the Euclidean distance. We can see from the plot that the statistic $Z_w(t)$ is  right skewed. The $p$-value approximations (\ref{eq:Zw_approx}) and (\ref{eq:Zw2_approx}) would then underestimate the true tail probabilities. On the other hand, $Z_\text{diff}(t)$ is right skewed for small values of $t$ and left skewed for large values of $t$, which would also affect the analytic $p$-value approximation derived based on asymptotic results.

Hence, we perform skewness correction to improve the analytical $p$-value approximations for finite sample sizes. As illustrated in Figure \ref{fig:skew}, the extent of the skewness depends on $t$, so we adopt a skewness correction approach discussed in \cite{chen2015graph} that does the correction up to different extents based on the amount of skewness at each value of $t$. In particular, the approach provides a better approximation to the marginal probability, $\boldP(Z_w(t) \in b+ dx /b)$, $\boldP(Z_\text{diff}(t) \in b+ dx /b)$, $\boldP(Z_w(t_1,t_2) \in b+ dx /b)$ and $\boldP(Z_\text{diff}(t_1,t_2) \in b+ dx /b)$,  through a cumulant generating function $\psi (\theta) = \log \boldE_P(e^{\theta z})$. By applying a change of measure $dQ_{\theta} = e^{\theta Z-\psi(\theta)}dP$, we can approximate the marginal probability by
\[ \frac{1}{\sqrt{2\pi (1 + \gamma \theta_b)}}\exp(-\theta_b b-x \theta_b/b + \theta^2_b(1+\gamma \theta_b/3)/2), \] 
where $\theta_b$ is chosen such that $\dot{\psi}(\theta_b) = b.$ By a third Taylor approximation, we get
$\theta_b \approx (-1 + \sqrt{1+2\gamma b})/\gamma, $
where $\gamma: = \boldE_\boldP(Z^3)$.

Notice that $ \boldE(Z^3_w(t_1,t_2)) = \boldE(Z^3_w(t_2 - t_1))$ and 
$ \boldE(Z^3_\text{diff}(t_1,t_2))=\boldE(Z^3_\text{diff}(t_2 - t_1)) $.  Let  $\gamma_w(t) = \boldE(Z^3_w(t))$ and $\gamma_\text{diff}(t) = \boldE(Z^3_\text{diff}(t))$.
The $p$-value approximations, after correcting for skewness, are
\begin{align}
\label{eq:Zw_pvalue_skew} 
 \boldP & \left(\max_{n_0 \le t \le n_1} Z_w(t) > b \right) \\ 
 & \approx b \phi(b) \int_{\tfrac{n_0}{n}}^{\tfrac{n_1}{n}} K_w(nx)h_w(n,x) \nu (b \sqrt{2h_w(n,x)/n})dx,  \nonumber \\
 \boldP &  \left(\max_{n_0 \le t_2 -t_1 \le n_1} Z_w(t_1,t_2) > b \right) \label{eq:Zw2_pvalue_skew} \\ 
&  \approx b^3 \phi(b) \int_{\tfrac{n_0}{n}}^{\tfrac{n_1}{n}} K_w(nx) \left( h_w(n,x) \nu (b \sqrt{2h_w(n,x)/n}) \right ) ^2 (1-x)dx \nonumber
\end{align}
where $ K_w(t) = \frac{\exp \left( \frac{1}{2} (b - \hat{\theta}_{b,w}(t))^2 + \frac{1}{6} \gamma_w(t) \hat{\theta}_{b,w}(t)^3 \right )}{\sqrt{1+\gamma_w(t)\hat{\theta}_{b,w}(t)}}$
with $ \hat{\theta}_{b,w}(t) = \frac{-1 + \sqrt{1+2 \gamma_w(t)b}}{\gamma_w(t)}, $
and
\allowdisplaybreaks
\begin{align}
 & \boldP \left(\max_{n_0 \le t \le n_1} Z_\text{diff}(t) > b \right) \label{eq:Zdiff_pvalue_skew} \\
    \approx\, &\, b \phi(b) \int_{\tfrac{n_0}{n}}^{\tfrac{n_1}{n}} K_\text{diff}(nx)h_\text{diff}(n,x) \nu (b \sqrt{2h_\text{diff}(n,x)/n})dx,   \nonumber \\
 & \boldP \left(\max_{n_0 \le t_2 -t_1 \le n_1} Z_\text{diff}(t_1,t_2) > b \right)  \label{eq:Zdiff2_pvalue_skew} \\
  \approx\, &\, b^3 \phi(b) \int_{\tfrac{n_0}{n}}^{\tfrac{n_1}{n}} K_\text{diff}(nx) \left( h_\text{diff}(n,x) \nu (b \sqrt{2h_\text{diff}(n,x)/n}) \right ) ^2 (1-x)dx, \nonumber 
\end{align}
where $ K_\text{diff}(t) = \frac{\exp \left( \frac{1}{2} (b - \hat{\theta}_{b,\text{diff}}(t))^2 + \frac{1}{6} \gamma_\text{diff}(t) \hat{\theta}_{b,\text{diff}}(t)^3 \right )}{\sqrt{1+\gamma_\text{diff}(t)\hat{\theta}_{b,\text{diff}}(t)}}$
with $  \hat{\theta}_{b,\text{diff}}(t) = \frac{-1 + \sqrt{1+2 \gamma_\text{diff}(t)b}}{\gamma_\text{diff}(t)}. $ 

The only unknown quantities in the above expressions are $\gamma_w(t)$ and $\gamma_\text{diff}(t)$.  
Since 
\begin{align*}
 \boldE[Z^3_w(t)] & = \frac{\boldE(R^3_w(t))-3\boldE(R_w(t))\text{\bf{Var}}(R_w(t)) - \boldE^3(R_w(t))}{(\text{\bf{Var}}(R_w(t)))^{3/2}}, \\
\boldE[Z^3_\text{diff}(t)] & = \frac{\boldE(R^3_\text{diff}(t))-3\boldE(R_\text{diff}(t))\text{\bf{Var}}(R_\text{diff}(t)) - \boldE^3(R_\text{diff}(t))}{(\text{\bf{Var}}(R_\text{diff}(t)))^{3/2}},
\end{align*}
and the analytic expressions for the expectation and variance of $R_w(t)$ 
and $R_\text{diff}(t)$ can be found in Section \ref{sec:3}, we only need to figure out the analytic expressions of $\boldE(R^3_w(t))$ and $\boldE(R^3_w(t_1,t_2))$. The exact analytic expressions of $\boldE(R^3_w(t))$ and $\boldE(R^3_w(t_1,t_2))$ are quite long and they are provided in Appendix \ref{app:extra}. 

\begin{remark}
When the marginal distribution is highly left-skewed, it is possible that the third moment of the test statistic, $\gamma(t)$,  is too small for $1+2\gamma(t)b$ to be positive. In order to obtain a better approximation to $\theta_{b}$, higher moments are needed. However, since this problem usually occurs when $t/n$ is close to $0$ or $1$, we apply a heuristic fix discussed in \cite{chen2015graph} that  extrapolates $\hat{\theta}$ by using its values outside the problematic region. 
\end{remark}

\begin{remark} \label{rmk: S_skew} Skewness corrected $p$-value approximations for \\$\max_{n_0 \le t \le n_1} S(t) = \max_{0\leq w\leq 2\pi} \max_{n_0 \le t \le n_1}  (Z_w(t) \sin(w) + Z_{\text{diff}}(t) \cos(w))$  can be derived by jointly correcting for the marginal probabilities of  $Z_w(t)$ and $Z_\text{diff}(t)$.
After correcting for skewness, the integrand in \eqref{eq:S_approx} becomes $$K_S(x, \omega)u(x,\omega) \nu(\sqrt{2bu(x,\omega)/n}),$$ where \\
$ K_S(t,\omega) = \tfrac{\exp \left( \frac{1}{2} \left((\sqrt{b}\cos(\omega) - \hat{\theta}_{b,1}(t))^2 + (\sqrt{b}\sin(\omega) - \hat{\theta}_{b,2}(t))^2\right) + \frac{1}{6}(\gamma_1(t)\hat{\theta}_{b,1}(t)^3 + \gamma_2(t)\hat{\theta}_{b,2}(t)^3) \right)}{\sqrt{(1+\gamma_1(t)\hat{\theta}_{b,1})(1+\gamma_2(t)\hat{\theta}_{b,2})}}$
with $\gamma_1(t) = \boldE[Z^3_1(t)],  \hat{\theta}_{b,1}(t,\omega) = \frac{-1 + \sqrt{1+2 \gamma_1(t)\sqrt{b}\cos(\omega)}}{\gamma_1(t)} $, and $\gamma_2(t)$ and $\hat{\theta}_{b,2}(t,\omega)$ defined similarly.   However, this integrand could easily be non-finite in each quadrant in terms of $w$, and the method relies heavily on extrapolation. We thus do not perform skewness correction on $S(t)$. 
\end{remark}

\subsection{Checking p-value approximations for finite samples}
\label{sec:4.4}

Here, we check how the $p$-value approximations based on asymptotic results directly and with skewness correction work for finite samples. To do so, we compare the critical values obtained from (\ref{eq:S_approx}), (\ref{eq:Zw_approx}), (\ref{eq:M_approx}), (\ref{eq:Zw_pvalue_skew}), and (\ref{eq:Zdiff_pvalue_skew}) to the critical values obtained from doing 10,000 permutations directly, under various simulation settings. We here focus on the single change-point alternative here. For the changed interval alternative, the results are similar and details can be found in Supplement \ref{app:CV}. 

In each simulation, sequences of length 1,000 were generated from a given distribution $F_0$ in $\mathbb{R}^d$. We considered three distributions (multivariate normal, multivariate $t$ with $5$ degrees of freedom, and multivariate log-normal) under various dimensions ($d=10$, $d=100$, and $d=1000$). Here, we present the results only for multivariate normal with $d=10$ (denoted by (C1) in Tables \ref{table:S_CV05_reduced}, \ref{table:Rw_CV05_reduced}, and \ref{table:M_CV05_reduced}), multivariate $t_5$ with $d=100$ (denoted by (C2)), and multivariate log-normal with $d=1000$ (denoted by (C3)). The complete tables showing all three distributions under these three dimensions with more cases are in Supplement  \ref{app:CV}. The analytical approximations depend on constraints on the sequence in which the change-point is searched over ($n_0$ and $n_1$). To make things simple, we let $n_1 = n - n_0$. 

Since the asymptotic $p$-value approximations (without skewness correction) do not depend on $G$, the critical value  is determined by $n$, $n_0$, and $n_1$ only (here, $n_1$ is set to be $n-n_0$). 
The first table of Tables \ref{table:S_CV05_reduced}, \ref{table:Rw_CV05_reduced}, and \ref{table:M_CV05_reduced} labeled `A1' presents the analytical critical values without skewness correction. 
On the other hand, the skewness corrected $p$-value approximations and permutation $p$-values depend on certain characteristics of the structure of the graph $G$.  In this simulation, the MST is used. As the structure of MST depends on the observations, the critical value vary by simulation runs. We show results for 2 randomly simulated sequences in each setting. Two characteristics of the graph are also reported: the sum of squared node degrees ($\sum_i|G_i|^2$) and the maximum node degree ($d_{\max}$). These quantities give some intuitions on the size and density of the hubs in the graph. The skewness corrected critical values are presented in Tables \ref{table:Rw_CV05_reduced} and \ref{table:M_CV05_reduced} under the column `A2'. The column `Per' denotes critical values obtained through 10,000 random permutations directly.

\begin{table}[h]
\centering 
 \caption{Critical values for the single change-point scan statistic $\max_{n_0 \le t \le n_1} S(t)$ based on MST at 0.05 significance level. n = 1000.}
  \label{table:S_CV05_reduced}
\begin{tabular}{|c|c|c|c|c|}
\hline  
\hline
& $n_0 = 100$ & $n_0 = 75$ & $n_0 = 50$ & $n_0 = 25$\\
\hline
A1  & 13.10 & 13.38 & 13.70 & 14.11 \\
\hline 
\end{tabular} 

\vspace{2mm}

\begin{tabular}{c|c|c|c|c|cc} 
\hline  
\hline
& \multicolumn{4}{c}{Critical Values} & \multicolumn{2}{|c}{Graph} \\
\cline{2-7}
& \multicolumn{1}{|c|}{$n_0 = 100$} &  \multicolumn{1}{|c|}{$n_0 = 75$} &  \multicolumn{1}{|c|}{$n_0 = 50$} &  \multicolumn{1}{|c|}{$n_0 = 25$} \\
\cline{2-5}
& Per &  Per & Per & Per & $\sum|G_i|^2$ & $d_{\max}$ \\
\hline 
\hline
\multirow{2}{0.7cm}{(C1)}
& 12.87 & 13.29 & 14.04 & 15.17 & 5394 &  8\\
& 13.02 & 13.42 & 13.71 & 15.65 & 5368 &  8\\
\hline
\multirow{2}{0.7cm}{(C2)}
& 13.47 & 14.20 & 15.48 & 17.81 & 14302 & 42 \\
& 13.32 & 13.77 & 14.96 & 17.11 & 12424 & 39 \\
\hline
\multirow{2}{0.7cm}{(C3)}
& 14.50 & 15.83 & 18.14 & 21.96 &  46876 &  83\\
& 16.12 & 18.38 & 22.00 & 29.07 & 106524 & 208\\
\hline
\hline 
\end{tabular} 
\end{table} 

\begin{table}[h]
\centering 
 \caption{Critical values for the single change-point scan statistic $\max_{n_0 \le t \le n_1} Z_w(t)$ based on MST at 0.05 significance level. n = 1000.}
  \label{table:Rw_CV05_reduced}
\begin{tabular}{|c|c|c|c|c|}
\hline  
\hline
& $n_0 = 100$ & $n_0 = 75$ & $n_0 = 50$ & $n_0 = 25$\\
\hline
A1  & 2.98 & 3.02 &  3.08 & 3.14 \\
\hline 
\end{tabular} 

\vspace{2mm}

\begin{tabular}{c|cc|cc|cc|cc|cc} 
\hline  
\hline
& \multicolumn{8}{c}{Critical Values} & \multicolumn{2}{|c}{Graph} \\
\cline{2-11}
& \multicolumn{2}{|c|}{$n_0 = 100$} &  \multicolumn{2}{|c|}{$n_0 = 75$} &  \multicolumn{2}{|c|}{$n_0 = 50$} &  \multicolumn{2}{|c|}{$n_0 = 25$} \\
\cline{2-9}
& A2 & Per &  A2 &Per & A2 &Per &  A2 &Per &  $\sum|G_i|^2$ & $d_{\max}$ \\
\hline 
\hline
\multirow{2}{0.7cm}{(C1)}
& 3.05 & 3.02 & 3.12 & 3.11 & 3.22 & 3.22 & 3.4 & 3.48 & 5518 & 10\\
& 3.05 & 3.05 & 3.12 & 3.14 & 3.22 & 3.25 & 3.4 & 3.45 & 5442 &  8\\
\hline
\multirow{2}{0.7cm}{(C2)}
& 3.05 & 3.04 & 3.12 & 3.15 & 3.22 & 3.31 & 3.39 & 3.62 & 14302 & 42 \\
& 3.05 & 3.06 & 3.12 & 3.13 & 3.22 & 3.29 & 3.39 & 3.54 & 12424 & 39 \\
\hline
\multirow{2}{0.7cm}{(C3)}
& 3.04 & 3.11 & 3.11 & 3.25 & 3.21 & 3.37 & 3.38 & 3.82 &  46876 &  83 \\
& 3.03 & 3.20 & 3.10 & 3.40 & 3.19 & 3.61 & 3.35 & 3.99 & 106524 & 208 \\
\hline
\hline 
\end{tabular} 
\end{table} 

\begin{table}[h]
\centering 
 \caption{Critical values for the single change-point scan statistic $\max_{n_0 \le t \le n_1} M(t)$ based on MST at 0.05 significance level. n = 1000.}
  \label{table:M_CV05_reduced}
\begin{tabular}{|c|c|c|c|c|}
\hline  
\hline
& $n_0 = 100$ & $n_0 = 75$ & $n_0 = 50$ & $n_0 = 25$\\
\hline
A1  & 3.23 & 3.27 &  3.32 & 3.38  \\
\hline 
\end{tabular} 

\vspace{2mm}

\begin{tabular}{p{0.7cm}|cc|cc|cc|cc|cc} 
\hline  
\hline
& \multicolumn{8}{c}{Critical Values} & \multicolumn{2}{|c}{Graph} \\
\cline{2-11}
& \multicolumn{2}{|c|}{$n_0 = 100$} &  \multicolumn{2}{|c|}{$n_0 = 75$} &  \multicolumn{2}{|c|}{$n_0 = 50$} &  \multicolumn{2}{|c|}{$n_0 = 25$} \\
\cline{2-9}
& A2 & Per &  A2 &Per & A2 &Per &  A2 &Per &  $\sum|G_i|^2$ & $d_{\max}$ \\
\hline 
\hline
\multirow{2}{0.7cm}{(C1)}
& 3.27 & 3.26 & 3.33 & 3.34 & 3.41 & 3.42 & 3.56 & 3.66 & 5518 & 10\\
& 3.27 & 3.29 & 3.33 & 3.34 & 3.41 & 3.44 & 3.56 & 3.67 & 5442 &  8\\
\hline
\multirow{2}{0.7cm}{(C2)}
& 3.30 & 3.33 & 3.38 & 3.44 & 3.48 & 3.55 & 3.67 & 3.89 & 14302 & 42 \\
& 3.29 & 3.31 & 3.36 & 3.40 & 3.46 & 3.54 & 3.64 & 3.85 & 12424 & 39 \\
\hline
\multirow{2}{0.7cm}{(C3)}
& 3.33 & 3.34 & 3.41 & 3.49 & 3.53 & 3.69 & 3.74 & 4.22 &  46876 &  83 \\
& 3.39 & 3.51 & 3.49 & 3.75 & 3.63 & 4.06 & 3.88 & 4.58 & 106524 & 208 \\
\hline
\hline

\end{tabular} 
\end{table}

We first focus on the results of the generalized edge-count test statistic $\max S(t)$. Since we do not perform skewness correction for $S(t)$, Table \ref{table:S_CV05_reduced} compares these analytical critical values (A1) with the critical values obtained from doing 10,000 permutations (Per). The main factors that influence the approximation accuracy of the analytical critical values are the minimum window size ($n_0$) and the structure of the graph. 
We see that, when the graph is relatively flat (such as in (C1) that the largest degree in the graph is relatively small), the asymptotic $p$-value approximation is doing reasonably well when $n_0\geq 50$.  As the graph becomes to have larger and larger hubs, $n_0$ needs to be larger to achieve a similar degree of accurancy. 



Table \ref{table:Rw_CV05_reduced} shows the results for $\max Z_w(t)$. Similarly to $S(t)$, as window size decreases and/or the maximum degree in the graph increases, the analytical critical values become less precise. However, the skewness corrected critical values perform much better than the critical values without skewness correction.  Under (C1), the maximum degree of the graph is in general small and the skewness-corrected $p$-value approximations are doing reasonably well for $n_0$ as low as 25.   When the maximum degree of the graph is less than 50, the skewness-corrected $p$-value approximations are doing quite well for $n_0\geq 50$ and not bad for $n_0=25$.  
For even larger maximum degree scenarios, the skewness-corrected $p$-value approximations are somewhat less conservative for $n_0\leq 100$ but the discrepancy is not that bad.


Table \ref{table:M_CV05_reduced} shows the results for $\max M(t)$.  The pattern is somewhat similar to that for $\max Z_w(t)$ with the skewness-corrected $p$-value approximations for $\max M(t)$ slightly more tolerant for hubs.   When the dimension is not too high ((C1) and (C2)), the maximum degree is less than 50, and the skewness-corrected $p$-value approximations are working very well when $n_0\geq 50$.  When the maximum degree is large (C3), the skewness-corrected $p$-value approximations are still doing pretty well for $n_0\geq 75$ in general. 


Overall, we see that the asymptotic critical values are on the right scale and are enough for detecting big changes. However, if one would like to have more accurate critical values, the skewness correction versions are recommended.  When this is needed, it would be good to first check the structure of the graph, such as its maximum degree, so that we have a better idea on how well the critical values are.

\section{Performance analysis} \label{sec:5} Here, we examine the performance of the three new test statistics under more settings through simulation studies. Since the proposed tests do not require the data to be from any specific distribution family, there are many possible alternatives. To have a good idea of the performance of the proposed tests, we examine the Gaussian data ($\mathbf{y_i}\sim N_d(\mu, \Sigma)$) where likelihood-based methods are available. We also checked other distributions to check the robustness of the tests in terms of the underlying distribution and these tables can be found in Supplement \ref{sec:power}. 

Under the Gaussian setting, if one assumes that, at the change-point, only the mean ($\mathbf{\mu}$) may change, the scan statistic over Hotelling's T$^2$ statistics can be used: $\max_{n_0\leq t\leq n_1} HT(t), \text{ with }HT(t) = \frac{t(n-t)}{n}(\bar{\mathbf{y}}_t - \bar{\mathbf{y}}^*_t)^T \tilde{\Sigma_t}^{-1}(\bar{\mathbf{y}}_t - \bar{\mathbf{y}}^*_t)$ where
$ \bar{\mathbf{y}}_t = \sum_{i=1}^t \mathbf{y}_i/t, \,  \bar{\mathbf{y}}^*_t = \sum_{i=t+1}^n \mathbf{y}_i/(n-t),$ and 
$\tilde{\Sigma}_t = (\sum_{i=1}^t (\mathbf{y}_i - \bar{\mathbf{y}}_t)(\mathbf{y}_i - \bar{\mathbf{y}}_t)^T + \sum_{i=t+1}^n(\mathbf{y}_i - \bar{\mathbf{y}}^*_t)(\mathbf{y}_i - \bar{\mathbf{y}}^*_t)^T)/(n-2).$
 If the variance may also change at the change-point, the scan statistic over the generalized likelihood ratio statistic can be used:
 $\max_{n_0\leq t\leq n_1} GLR(t)$ with $GLR(t) = n \log |\hat{\Sigma}_n| - t \log |\hat{\Sigma}_t| - (n-t)\log|\hat{\Sigma}^*_t|,$
$\text{ where } \hat{\Sigma}_t = \frac{\sum_{i=1}^n(\mathbf{y}_i - \bar{\mathbf{y}}_t)(\mathbf{y}_i - \bar{\mathbf{y}}_t)^T}{t}$ and $ \hat{\Sigma}^*_t = \frac{\sum_{i=t+1}^n (\mathbf{y}_i - \bar{\mathbf{y}}^*_t)(\mathbf{y}_i - \bar{\mathbf{y}}^*_t)^T}{n-t}. $


In each simulation, we generated a sequence of $n=200$ observations for various dimensions $d$ with $\by_1,\dots, \by_\tau \overset{iid}{\sim} F_0$ and $\by_{\tau+1}, \dots, \by_n\overset{iid}{\sim} F_1$.  Here, $\tau$ is the change-point.
 When there is a mean difference, we use $\Delta$ to denote the Euclidean distance of the means of $F_0$ and $F_1$.  When there is a variance difference,  to make the change less significant, only the first $[d/5]$ of the diagonal elements of the covariance matrix differ with a multiple of $\sigma$, and the rest are unchanged.

For the proposed methods, we also expand our study to denser graphs, the $k$-MST, which is the union of the $1st, \hdots, k$th MSTs, where the 1st MST is the MST and the $i$th MST ($i>1$) is the spanning tree with the sum of the distances of the edges in the tree minimized subject to the constraint that it does not use any of the edge in the 1st, \dots, $(i-1)$th MST(s).
Simulation studies show that the edge-count two-sample tests have higher power when the graph is slightly denser as it contains more similarity information. However, the optimal choice of $k$ is still an open question. \cite{chen2017new} recommend to use $5$-MST for the generalized edge-count two-sample test. In the following simulation settings, for simplicity, we set the graph to be the $5$-MST constructed using Euclidean distance. 

The performance of six methods are compared: two methods based on normal theory ($\max HT(t)$, $\max \text{GLR}(t)$), the method in \cite{chen2015graph} ($\max Z(t)$), and three new tests ($\max S(t)$, $\max R_w(t)$, $\max M(t)$). The estimated power is calculated as the number of trials, out of 100, that the null hypothesis is rejected at 0.05 level for each of these methods, with $p$-values determined by 10,000 permutation runs for fairness in comparison.  To examine the accuracy of the estimated change-point, the number of trials where the estimated change-point is within 20 from the true change-point is provided in parentheses.  Under each setting, the specific alternative is chosen so that the tests have moderate power to be comparable.  The best one for each scenario is made bold.  In the following, we use `HT' to refer to the scan statistic over the Hotelling's $T^2$ statistic and use `GLR' to refer to the scan statistic over the generalized likelihood ratio statistic.

\begin{table}[!htp] \centering 
\caption{Multivariate Gaussian data, mean difference, $\tau$ at center }
\begin{tabular}{p{1cm}|ccccccccccccc}
\hline
\hline
  d      & 10  & 50 & 100 & 150 & 175 &  500 & 2000\\
$\Delta$ & 0.8 & 1  & 1.2 & 1.6 & 2 & 2.5    & 3.4 \\
\hline
\hline
$HT$ & \textbf{91} \textbf{(83)}  & \textbf{82} \textbf{(72)} &  \textbf{72} \textbf{(60)}   &  65 51& 38 (26) & -   & -  \\
\hline
GLR & 22 (9) & 4 (0)  & - & - & -   & - & - \\
\hline
$Z$& 51 (46)& 50 (44) & 50 \textbf{(46)}& \textbf{84} (80)& \textbf{91} \textbf{(88)}& \textbf{91} \textbf{(91)}&  87 \textbf{(84)} \\
\hline
$Z_w$ & 39 (28)  & 45 (31)& 52 (32)& 78 (66)& 89 (79)& \textbf{91} \textbf{(85)}& \textbf{88} (78) \\
\hline
$S$ & 32 (21)& 33 (23) & 37 (23)& 68 (55)& 80 (69) & 84 (80)& 81 (71)\\
\hline
$M$ & 35 (26)& 36 (26)& 41 (25)  & 74 (63)& 86 (76)& 87 (82)& 86 (75)\\
\hline
\hline
\end{tabular}
\label{table:l_100} 
\vspace{1mm}
\caption{Multivariate Gaussian data, mean difference, $\tau$ at three quarters }
\begin{tabular}{p{1cm}|ccccccccccccc}
\hline
\hline
  d      & 10  & 50 & 100 & 150 & 175 &  500  & 2000\\
$\Delta$ & 0.8 & 1  & 1.2 & 1.6 & 2 & 2.5  & 3.4 \\
\hline
\hline
$HT$ & \textbf{75} \textbf{(63)} & \textbf{70} \textbf{(66)}& \textbf{48} \textbf{(38)}& 40 (28)& 34 (30)&  - &  - \\
\hline
GLR & 16 (8)& 12 (8) &  - &  - &  - &  - &  -  \\
\hline
$Z$ & 25 (15)& 14 (5)& 17 (7)& 17 (6)& 42 (12)& 37 (14)&  30 (11)\\
\hline
$Z_w$ & 29 (23)& 25 (18)& 31 (20)& \textbf{52} \textbf{(42)}& \textbf{63}  \textbf{(55)}& \textbf{67} \textbf{(55)}&  \textbf{68} \textbf{(62)}\\
\hline
$S$ & 25 (16)& 17 (10)& 25 (17)& 35 (29) & 50 (46)& 49 (39)& 48 (44)\\
\hline
$M$ & 25 (21)& 20 (15)& 29 (17)& 41 (32)& 53 (48)& 62 (51)& 58 (53)\\
 \hline\hline
\end{tabular}
\label{table:l_150} 
\vspace{1mm}
\vspace{1mm}
\caption{Multivariate Gaussian data, mean and scale difference, $\tau$ at center }
\begin{tabular}{c|ccccccccccccccc}
\hline
\hline
  d      & 10  & 50   & 100 & 150 & 175  &  500 & 2000\\
$\Delta$ & 0.6 & 1    & 1.2 & 1.2 & 1.05 & 1    & 1 \\
$\sigma$ & 1.3 & 1.3  & 1.1 & 1.1 & 1.1  & 1.1  & 1.05 \\
\hline
\hline
$HT$ & \textbf{49} \textbf{(34)}&  73 (60)  &  \textbf{65} \textbf{(53)}& 30 (16) & 15 (5) & - & -  \\
\hline
GLR & 26 (17)&  12 (0) &  -  &  - & - & - & -\\
\hline
$Z$ & 38  (28) &  \textbf{79}(66) & 62 (52)& 47\textbf{(34)} & 39 (29)& 55 (33) &  54 (18)\\
\hline
$Z_w$ & 30 (14) & 62 (55)&  55  (43)& 38 (28)& 29 (21)& 15 (4) & 18 (5) \\
\hline
$S$ & 30 (17)& \textbf{79} \textbf{(70)} &  49 (37)& \textbf{48} (30)& \textbf{44} \textbf{(36)}& 66 (42) &  69 (44) \\
\hline
$M$ & 29 (12)& 76 (67)& 52 (40) & 51 (29)& 43 (34)& \textbf{69} \textbf{(50)}& \textbf{74} \textbf{(51)} \\
\hline \hline
\end{tabular}
\label{table:ls_100} 
\vspace{2mm}
\caption{Multivariate Gaussian data, mean and scale difference, $\tau$ at three quarters }
\begin{tabular}{p{1cm}|cccccccccccccccc}
\hline
\hline
  d      & 10  & 50    & 100 & 150  & 175  &  500 &  2000\\
$\Delta$ & 0.6 & 1     & 1.2 & 1.1  & 1.05 & 0.9  & 0.95 \\
$\sigma$ & 1.3 & 1.15  & 0.9 & 0.85 & 0.85 & 0.8  & 0.6 \\
\hline
\hline
$HT$ & \textbf{37} \textbf{(28)}& \textbf{63} \textbf{(53)}& \textbf{43} \textbf{(36)}& 11 (4)&  9 (3)& - & -  \\
\hline
GLR & 17 (8) &  8 (5)&  - &  - & - & - & - \\
\hline
$Z$ & 37 (23)&   34 (20)  &  16 ( 0)&  21 ( 0) &  18 ( 0)& 13 ( 0) &  15 ( 0)\\
\hline
$Z_w$ & 23 (16)&   21 (12)&  34 (29)&  36 (22)&  34 (23)& 15 ( 9) &  4 ( 1)\\
\hline
$S$ & 25 (18)&  22 (12)&   36 (29)&  44 (34)&  \textbf{56} \textbf{(45)} & 54 (48)&\textbf{57} (52)\\
\hline
$M$  & 23  (15)  &  19 (10) & 34 (27)& \textbf{48} \textbf{(38)} & 53 (41)&  \textbf{58} \textbf{(52)}&  \textbf{57} \textbf{(54)}\\
 \hline \hline
\end{tabular}
\label{table:ls_150} 
\end{table}

Tables \ref{table:l_100} - \ref{table:ls_150} show results for multivariate Gaussian data under various alternatives.  When there is a mean change only (Tables \ref{table:l_100} and \ref{table:l_150}), we see that in general HT outperforms all other methods in low to moderate dimensions.  As dimension becomes larger, the graph-based tests take over.  When the location change occurs in the middle of the sequence, the scan statistic $Z(t)$ from \cite{chen2015graph} outperforms all other tests as dimension increases (Table \ref{table:l_100}) and the advantage of $Z_w(t)$ becomes evident (Table \ref{table:l_150}).  


Results for scale change only can be found in Supplement \ref{sec:power}. Under this setting, when dimension is low GLR dominates in power. But starting at $d=20$, the graph-based methods exceed GLR in power and $S(t)$ and $M(t)$ have much higher power among the graph-based methods. More details can be found in Supplement \ref{sec:power}. 

When there is both location and scale change (Tables \ref{table:ls_100} and \ref{table:ls_150}), we see that when dimension is low, the parametric-based scan statistics dominate in power. As dimension increases, the new graph-based methods exceed $Z(t)$ and the parametric methods in power. Depending on the size of the change, the best graph-based method is different.
Generally, $M(t)$ seems to be most effective in detecting and estimating change-points for high dimension compared to the other graph-based test statistics. 


The overall pattern of the power tables show that when $d$ increases the graph-based statistics dominate the parametric tests. The new graph-based methods perform well under various scenarios. In general, $Z_w(t)$ dominates under the alternative of location change away from the center of the sequence whereas $M(t)$ and $S(t)$ dominate under the alternatives of change not only in location. Even under the scenario that is well-suited for the method in \cite{chen2015graph} (location change at the center of the sequence), the new graph-based methods perform at a comparable level to the old method.   Based on these results, if one is certain that the change is locational, the test based on $Z_w(t)$ is recommended; while for more general changes, the tests based on $S(t)$ and $M(t)$ are recommended.

\section{A real data example}
\label{sec:6}
We illustrate the new approaches on the yellow taxi trip records, which is publicly available on the NYC Taxi \& Limousine Commission (TLC) website (\url{http://www.nyc.gov/html/tlc/html/about/trip_record_data.shtml}). The trip records give information on the taxi pickup and drop-off date/times, longitude and latitude coordinates of pickup and drop-off locations, trip distances, fares, rate types, payments types, and driver-reported passenger counts. 

This dataset is very rich and many questions can be posed. Here, we illustrate the new approach in detecting changes in travel from the John F. Kennedy International Airport for the months October through December of 2015. For simplicity, the boundary of JFK airport was set to be $40.63$ to $40.66$ latitude and $-73.80$ to $-73.77$ longitude. 

For those trips that began with a pickup at JFK, we extract information on their longitude and latitude drop-off coordinates. Using longitude/latitude coordinates, we create a $30$ by $30$ grid of New York City and count the number of taxi drop-offs that fall within each cell, where each cell represents a longitude, latitude coordinate range. Then for each day, we have a $30$ by $30$ matrix such that each element represents the number of taxi drop-offs in each location. 

\begin{figure}[h]
  \begin{minipage}[b]{0.45\linewidth}
    \centering
    \includegraphics[width=\linewidth]{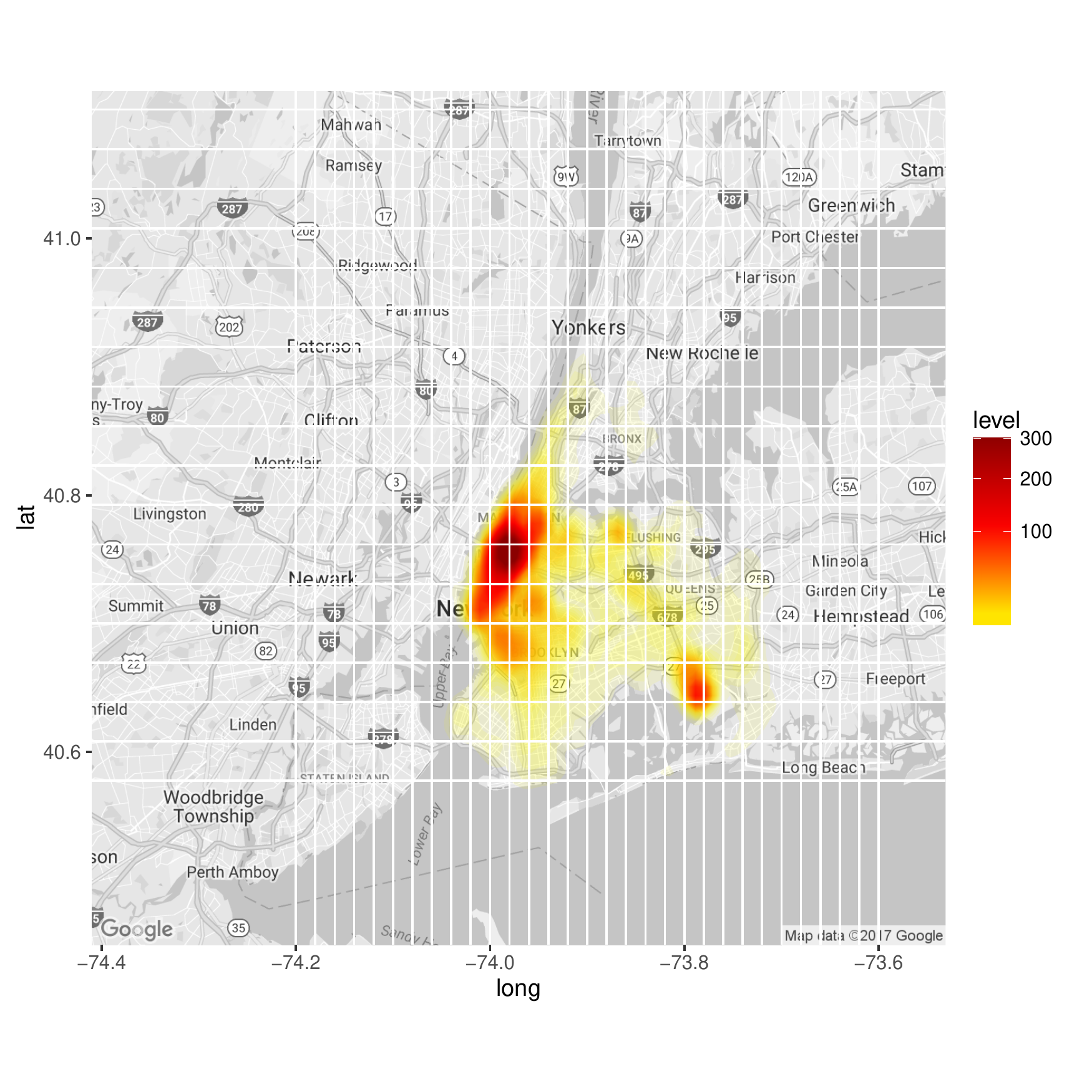} \\ (a) Oct. 15, 2015
    \vspace{3ex}
  \end{minipage}
  \begin{minipage}[b]{0.45\linewidth}
    \centering
    \includegraphics[width=\linewidth]{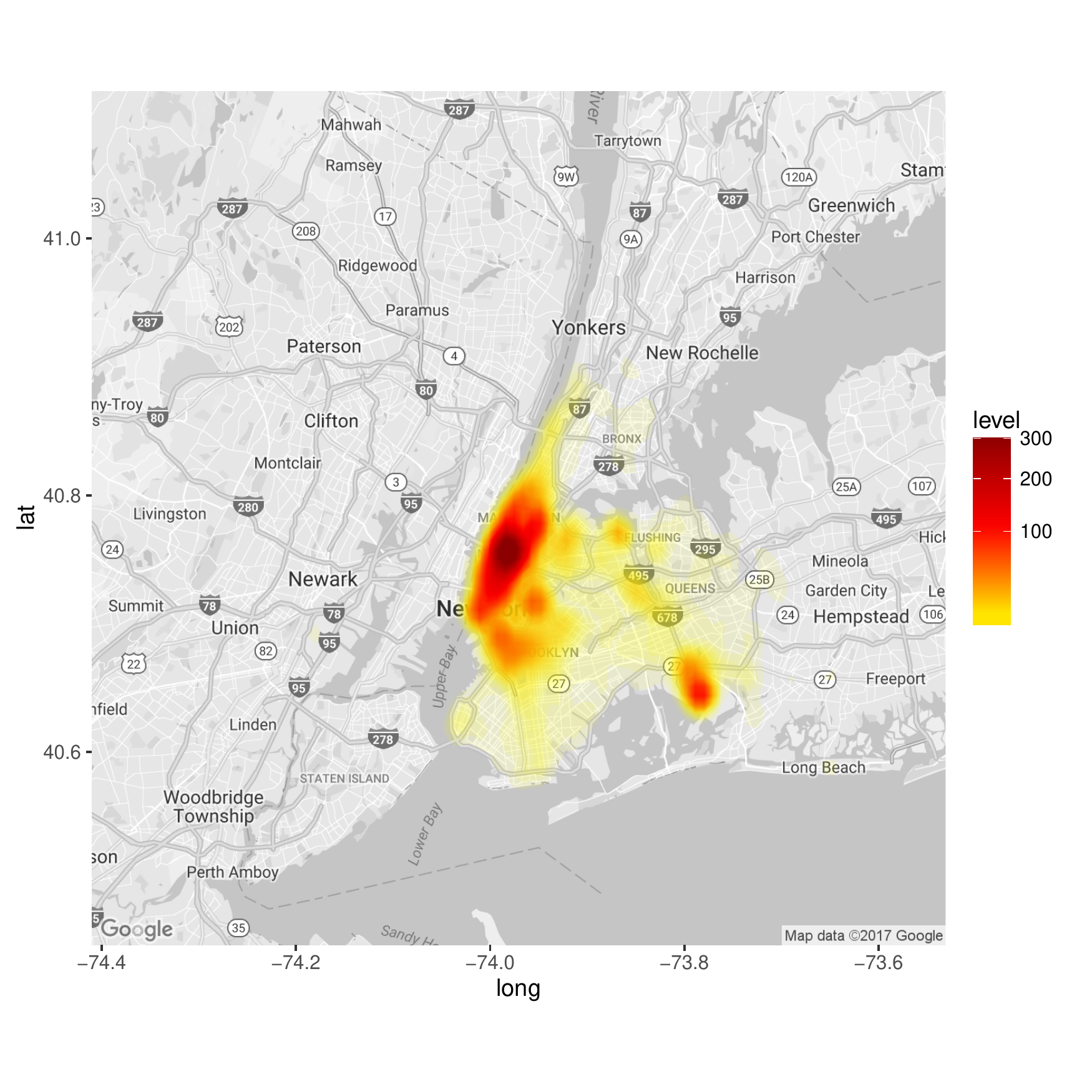} \\ (b) Nov. 20, 2015
    \vspace{3ex}
  \end{minipage} 
\vspace{-5mm}
  \caption{Density heatmap of taxi drop-offs for four randomly selected days.}
  \label{fig:heatmaps}
\end{figure}

The NYC taxi dataset is immense in both size and information. To better visualize the dataset, we plot heatmaps of the frequency of taxi drop-offs for a small area of New York City that cover the drop-off locations. Figure \ref{fig:heatmaps}, provides an illustration of the $30$ by $30$ grid we construct and two randomly selected days in our 3 month period: Oct. 15 and Nov. 20 for visualization. Heatmaps of additional days can be found in Supplement \ref{sec:NYfigures}.  The overall patterns are similar, but a more careful examination reveals there are some differences. To test whether differences are just by randomness or there is a significant change, we apply the three new approaches together with the method in \cite{chen2015graph}.

Let $A_i$ be the $30$ by $30$ matrix on day $i$. We denote $v_i$ to be the vector form of $A_i$, which is now $900$ by $1$. The $L_1$ norm is used to construct the MST graph representing similarity between days.

For the period of Oct. 1 through Dec. 31, the edge-count statistic $Z(t_1,t_2)$ reports 11/21/15 - 12/31/15 (Day 52 - 92) as the changed interval result. However, the new approaches all report the week right before Christmas, 12/18/15 - 12/25/15 (Day 79 - 86), as the changed interval (Table \ref{table:pickup}).
All these tests reject the null hypothesis of no change, with $p$-value $<0.001$.   

As there might be more than one changed interval, we further perform the tests on the period Oct. 1 through Dec. 17.  
During this time period, $Z(t_1,t_2)$ selects 10/27/15 - 12/17/15 (Day 27 - 78) as the changed interval. The new test statistics all report the week right before Thanksgiving, 11/20/15 - 11/27/15 (Day 51 - 58), as the changed interval.  All these tests reject the null hypothesis of no change as well, with $p$-value $<0.001$.

We further continued this process by performing the test on the period Oct. 1 through Nov. 20.  
The original edge-count test  $Z(t_1,t_2)$ reports a changed interval from 10/22/15 - 11/19/15 (Day 22 - 50).  It reject the null hypothesis of no change as well, with a small $p$-value (0.0017).  All three new tests report a changed interval of 11/16/15 - 11/19/15 (Day 47 - 50) but fail to reject the null hypothesis at the $0.01$ significance level.

\begin{table}[h] \centering
\caption{Changed interval results and corresponding p-values (reported in parentheses) for NYC taxi pickups from JFK.} \label{table:pickup}

\begin{tabular}{|c|c|c|c|c| } 
 \hline
Time period & $Z$ & $Z_w$ & $S$ & $M$ \\
\hline
\hline
10/1-12/31 & 11/21 - 12/31  &  12/18 - 12/25  & 12/18 - 12/25  & 12/18 - 12/25  \\ 
& ($<0.001$)     & ($<0.001$)   & ($<0.001$)   & ($<0.001$)  \\
\hline
10/1-12/17 & 10/27 - 12/17  &  11/20 - 11/27  & 11/20 - 11/27  & 11/20 - 11/27  \\ 
& ($0.0011$)     & ($<0.001$)   & ($<0.001$)   & ($<0.001$)  \\
\hline
10/1-11/20 & 10/22 - 11/19  &  11/16 - 11/19  & 11/16 - 11/19  & 11/16 - 11/19  \\ 
& ($0.0017$)   & ($0.0414$)   & ($0.0109$)   & ($0.0428$)  \\
\hline
\end{tabular} 
\end{table}
%
%

\begin{figure}[h]
    \centering
     \includegraphics[width=0.45\textwidth]{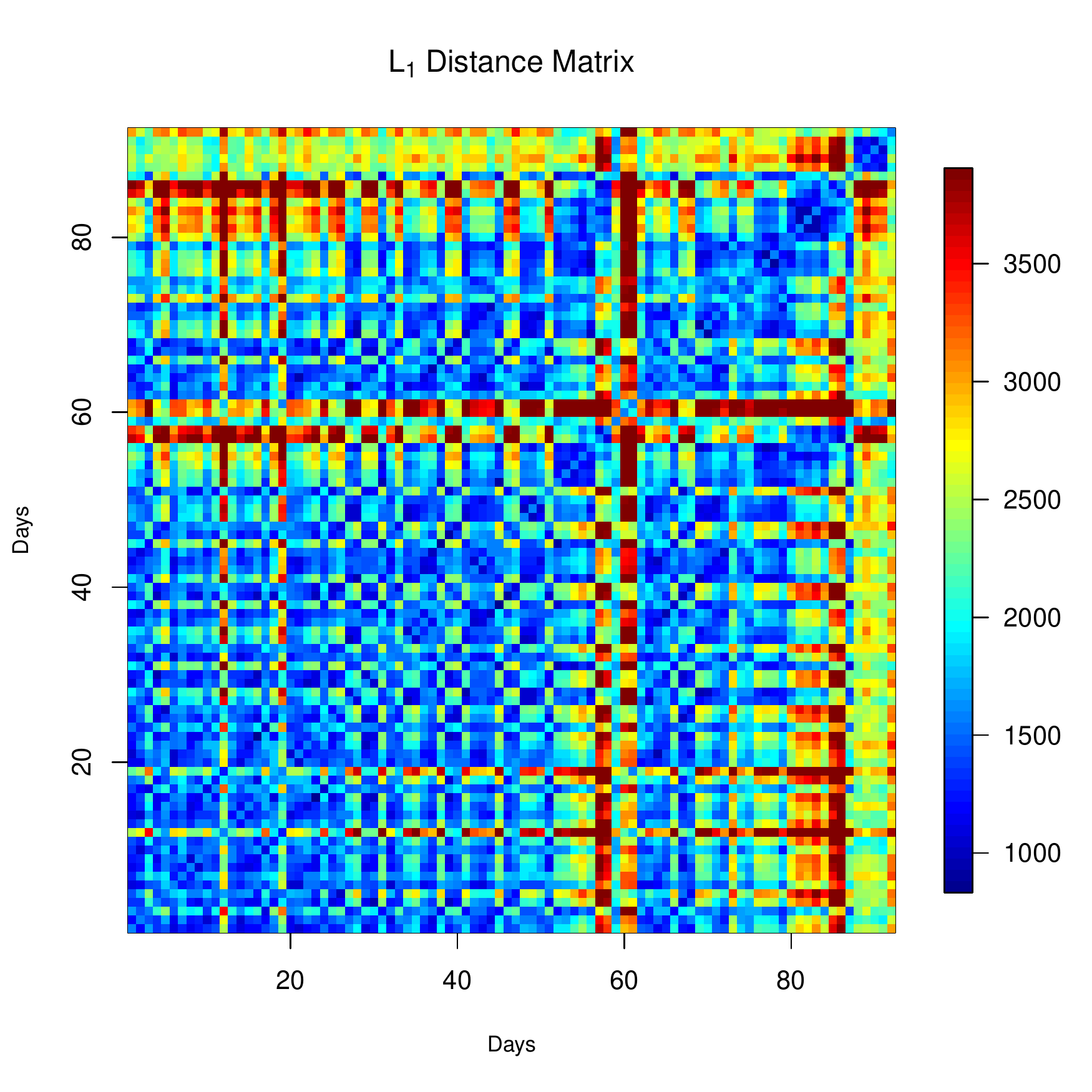} \ \ 
       \includegraphics[width=0.45\textwidth]{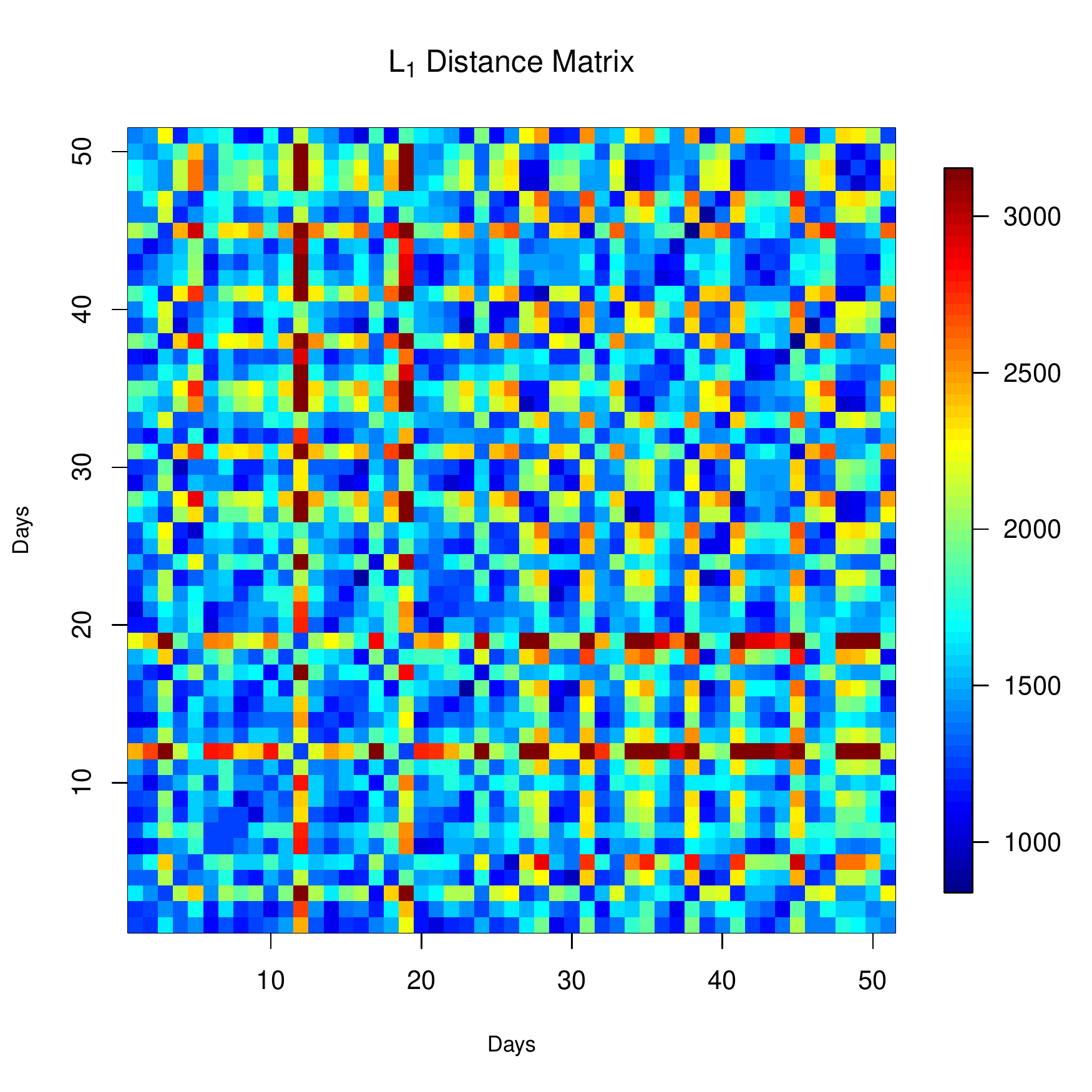} 
    \caption{Left panel: Heatmap of $L_1$ norm distance matrix of vector $v_i$ for $i = 1, \hdots 92$, corresponding to dates Oct. 1, 2015 - Dec. 31, 2015. Right panel: Heatmap of $L_1$ norm distance matrix of vector $v_i$ for $i = 1, \hdots 51$, corresponding to dates Oct. 1, 2015 - Nov. 20, 2015.}
    \label{fig:distance_heatmap}
\end{figure}

From the reported changed intervals, the results from the three new tests are more sensible -- the week right before Thanksgiving and the week right before Christmas.  To perform more sanity check, we plot the distance matrix of this whole period (Figure \ref{fig:distance_heatmap}, left panel).  It is evident that there is some change occurring around Day 60 and Day 80, matching with the results from the new tests.   On the other hand, the distance matrix for the first 51 days seems much more uniform (Figure \ref{fig:distance_heatmap}, right panel).

\section{Discussion and Conclusion}
\label{sec:8}

We propose new graph-based scan statistics for the testing and estimation of change-points that improve upon the framework proposed by \cite{chen2015graph}. Under various common scenarios, the new tests have improved power to detect changes and produce more precise estimates of the location of change-points.  

The new scan statistics are based on two basic processes, $Z_w(t)$ and $Z_\text{diff}(t)$, with the former sensitive to locational alternatives and the latter sensitive to scale alternatives.  These two basic processes rescaled by the length of the sequence -- $\{Z_w([nu]): 0<u<1\}$ and $\{Z_\text{diff}([nu]):0<u<1\}$ -- converge to independent Gaussian processes in finite dimensional distributions under some mild conditions of the graph.  The covariance functions of the limiting Gaussian processes do not depend on the graph, so the limiting processes are not affected by the distribution of the observations. 

Analytic $p$-value approximations based on limiting distributions (asymptotic $p$-value approximation) are derived for all new statistics and the skewness-corrected versions are derived for the weighted edge-count statistic and the max-type edge-count statistic.  The asymptotic $p$-value approximations provides a ballpark estimate of the $p$-value.  The skewness-corrected versions give more accurate approximations.  Based on simulation studies, even when the conditions for the graph in deriving the limiting distribution were violated, the analytic $p$-value formulas still give reasonable approximations.  A more detailed discussion on the conditions is in Supplement  \ref{sec:condition}. 


The performance of the new tests are examined under a number of settings.  Simulation results show that the weighted edge-count statistic is extremely useful when the change is locational and  the change-point not close to the center of the sequence.  When the change in the variance of the distribution is also of interest, the generalized edge-count statistic and the max-type edge-count statistic are recommended.   Together with the fact that the skewness-corrected $p$-value approximations can be easily obtained for the max-type edge-count statistic,  the test based on $M(t)$ is preferred to use. 

When the independence assumption is violated, instead of using the permutation null, we could do block permutation, i.e., the sequence is divided into blocks of size $b$ and the blocks are permuted.  In this way, the local structure in the sequence is retained.  All these test statistics can be modified accordingly to account for local dependence.  The detailed information is in 
Supplement \ref{sec:bp}.




\begin{supplement}
\sname{Supplement to 'Asymptotic Distribution-free Change-point Detection for Multivariate and Non-Euclidean Data'} 
\slink[url]{http://www.e-publications.org/ims/support/dowload/imsart-ims.zip}
\sdescription{The supplementary material contains the new test statistics for the changed-interval alternative, additional technical results and proofs, more illustrations of the data, additional power and analytical critical value tables, and further discussion on the conditions of the graph and the relationship between the new statistics, including an extension of the max-type statistic.  The table of contents of the supplement is listed below.}
\end{supplement}
\renewcommand\appendixname{Supplement}
\appendix
\section{Test statistics for changed interval alternative}\label{app:A}

%

\section{Exact analytic expressions for third moments}\label{app:extra}

\section{Proofs for Lemmas, Propositions and Theorems}\label{app:Proofs}

%
%
%

\section{Checking analytic $p$-value approximations}  \label{app:CV}

%

\section{Power tables} 
\label{sec:power}

\section{Additional illustrations of the NYC taxi dataset} 
\label{sec:NYfigures} 

\section{More discussions on the conditions}
\label{sec:condition}

%

\section{Relationship of the three test statistics and an extension to the max-type edge count test}
\label{sec:Mk}

\section{Block permutation for local dependency}
\label{sec:bp}

\section*{Acknowledgments}
Lynna Chu and Hao Chen are supported in part by NSF award DMS-1513653.

\bibliography{References}
\bibliographystyle{imsart-nameyear}

\end{document}